\documentclass[12pt]{article}

\usepackage[utf8]{inputenc}
\usepackage{graphicx}
\usepackage{float}
\usepackage{bbm}
\usepackage{amsmath}
\usepackage{braket}
\usepackage{amsfonts}
\usepackage{amssymb}
\usepackage{appendix}
\begin{document}

\begin{center}

\textbf{\Large Neutrino Invisible Decay at DUNE: a multi-channel analysis  
                } 

\vspace{30pt}
Anish Ghoshal$^{a,b}$, Alessio Giarnetti$^a$ and Davide Meloni$^a$
\vspace{16pt}

\textit{$^a$Dipartimento di Matematica e Fisica, 
Universit\`a di Roma Tre\\Via della Vasca Navale 84, 00146 Rome, Italy}\\
\vspace{16pt}
\textit{and}\\
\vspace{16pt}

\textit{$^b$
INFN,
Laboratori Nazionali di Frascati, C.P. 13, 00044 Frascati, Italy}\\
\vspace{16pt}

\end{center} 

\abstract
The hypothesis of the decay of neutrino mass eigenstates leads to a substantial modification of the appearance and disappearance probabilities of flavor eigenstates. We investigate the impact on the standard oscillation scenario caused by the decay of the heaviest mass eigenstate $\nu_3$ (with a mass $m_3$ and a mean life $\tau_3$) to a sterile state in DUNE. We find that the lower bound of $5.1 \times 10^{-11}~s/eV$ at 90\% CL on the decay parameter $\tau_3/m_3$ can be set if the Neutral Current data are included in the analysis, thus providing the best long-baseline expected limit so far. We also show that the $\nu_\tau$ appearance channel would give only a negligible contribution to the decay parameter constraints.
Our numerical results are corroborated by analytical formulae for the appearance and disappearance probabilities in vacuum (which is a useful approximation for the study of the invisible decay model) that we have developed up to the second order in the solar mass splitting  and to all orders in the decay factor $t/\tau_3$.

\section{Introduction}

Several neutrino experiments point to the fact that neutrinos have mass (at least two of them are non-vanishing) and they oscillate among three distinct flavors. 
However, the questions of the origin of neutrino masses, their nature (Dirac or Majorana) and the explanation of the mixing pattern (mixing angles and phases) from the first principles are still unanswered and most certainly call for new physics beyond the Standard Model.

The most important goals of the next generation neutrino oscillation experiments are the precise measurements of the leptonic CP-violating phase, the determination of the mass ordering of the neutrino states (normal or inverted hierarchy)
and the resolution of the $\theta_{23}$ octant degeneracy.
Furthermore, since we are entering the era of precise measurements in the leptonic sector, it is mandatory to study whether possible departures of the experimental data  
from the expected {\it standard} results can be ascribed to  new physics phenomena in the oscillations, such as those involving Non-Standard Neutrino Interactions (NSI) \cite{Roulet:1991sm,Guzzo:1991hi}, the presence of sterile neutrinos 
\cite{Maltoni:2007zf} and Lorentz invariance violations \cite{Barenboim:2017ewj} among others.

Another interesting form of non-standard physics not strictly related to neutrino oscillation is the possibility that neutrinos (or some of them) are unstable particles that can decay to lighter degrees of 
freedom \footnote{Massive neutrinos can undergo radiative decay to lighter neutrinos; however they are tightly constrained and will not be discussed here \cite{Mirizzi:2007jd}.} \cite{Smirnov,Lindner:2001th}. Such a possibility is contemplated in models where neutrinos couple to massless scalar fields, often called Majoron $S$ \cite{Gelmini:1983ea,Schechter:1981cv,Gelmini:1980re,Chikashige:1980ui} 
 via scalar ($g_s$) and pseudoscalar ($g_p$) couplings \cite{Barger:1999bg,Lindner:2001fx,Beacom:2002cb}:
\begin{equation}
\mathcal{L}_{\rm int}=\frac{(g_s)_{ij}}{2}\bar\nu_i\nu_jS+i\frac{(g_p)_{ij}}{2}\bar\nu_i\gamma_5\nu_j S~.
\end{equation}
The two terms lead to unstable neutrinos decaying through:
\begin{equation}
 \label{eq:decay}
\nu_i \rightarrow \nu + S\,,
\end{equation}
where $\nu_i$ ($i=1,2,3$) is a neutrino mass eigenstate with mass $m_{i}$, and the outgoing $\nu$ can be either an active and therefore observable neutrino state ({\it visible decay, VD}), or a {\it sterile} unknown neutrino state $\nu_s$ ({\it invisible decay, ID}). In this paper we will focus on the latter only. 

From the phenomenological point of view, the neutrino decay can be described by means of the {\it depletion factor}
\begin{equation}
D_i(t)=e^{- \frac{t}{\tau_i}}\, 
\label{dec}
\end{equation}
which, in the case of relativistic neutrinos, can be rewritten as

\begin{equation}
D_i(t)=e^{-\frac{m_i}{\tau_i} \frac{L}{E}}=e^{-\frac{1}{\beta_i} \frac{L}{E}}\, ,
\label{dec1}
\end{equation}
where  $E$ is the neutrino energy, $L$ is the experiment baseline and $\beta_i = \tau_i /m_i$ is the so-called {\it decay parameter}. As $D_i$'s depend on the $L/E$ ratio, we expect that oscillation probabilities will be affected by neutrino decays, especially when $\beta_i$ are small enough to have $\frac{1}{\beta_i} \frac{L}{E} \gg 1$.

The origin of the decay model analyzed here can be traced back to the 70s \cite{Bahcall:1972my}, when it was originally proposed to solve the solar neutrino anomaly through the decay of the mass eigenstate $\nu_2$.
However, neutrino decay alone could not explain the solar neutrino deficit and the hypothesis of flavor oscillations was invoked to explain the data \cite{Acker:1993sz}. This relegated  the $\nu_2$ decay, if present, to a sub-dominant process,  
as extensively analyzed in Refs.~\cite{Berezhiani:1991vk,Berezhiani:1992xg,Bandyopadhyay:2001ct,Bandyopadhyay:2002qg}.
From the solar neutrino oscillation data, the most stringent 90\% confidence level (CL) bound on the $\nu_2$ mass eigenstate decay parameter $\beta_2$ in the ID scheme is at 99\% CL:
\begin{equation}
 \beta_2 > 7\times 10^{-4} \;{\rm s/eV}\,,
\end{equation}
as obtained in \cite{Picoreti:2015ika,Berryman:2014qha,Huang:2018nxj}. 
Therefore, this decay of the $\nu_2$ eigenstate is strongly constrained and will not be considered here.

The decay of $\nu_3$, on the other hand, was proposed to explain the atmospheric neutrino problem but, just like in the case of the solar neutrino deficit, it could not provide a full explanation of the anomaly~\cite{Lipari:1999vh}.
For the ID hypothesis, long-baseline data from T2K and MINOS were able to set the following 90\% CL lower bounds \cite{Gomes:2014yua}: 
\begin{eqnarray}
\nonumber \beta_3 &>& 7.8 \times 10^{-13}  {\rm ~s/eV~(T2K)}
\\  &\label{bounbsb3}& 
\\\beta_3 &>& 2.8 \times 10^{-12} {\rm ~s/eV~(MINOS)}  \;.\nonumber
\end{eqnarray}
It is clear that the $\nu_3$ decay parameter is more difficult to constrain with the current data due to the (so far) limited statistics in long-baseline experiments \footnote{Another limit on the decay parameter set with long-baseline data has been obtained using the $\nu_\tau$ appearance events in OPERA, see \cite{Pagliaroli:2016zab}. }.
Note, however, that these results have been derived under the two-neutrino approximation and, therefore, a full three-neutrino analysis may loosen the bounds and significantly change the statistical relevance (and position in the parameter space) of possible best fits. 
Recently, following a three-neutrino approach, a new but even worse  90\% CL constraint on the neutrino decay lifetime has been obtained 
from the combination of T2K and NO$\nu$A data~\cite{Choubey:2018cfz}, at the level of
\begin{eqnarray}
\beta_3 > 1.5 \times 10^{-12} \;{\rm s/eV\qquad  T2K + NO\nu A}\,.
\end{eqnarray}
Finally, the best 99\% CL constraint has been set by the  combination of atmospheric and long-baseline data from Super-Kamiokande (SK), K2K and MINOS~\cite{GonzalezGarcia:2008ru}:
\begin{equation}
\label{bounbsb3_2}
\beta_3 > 9.3 \times 10^{-11} \;{\rm s/eV\qquad  SK + K2K + MINOS}\,.
\end{equation}
We want to remark that cosmological  observations \cite{Hannestad:2005ex,Escudero:2019gfk,Chacko:2019nej,Serpico:2007pt} can set limits on $\beta_3$ which are orders of magnitude larger than the ones predicted by long-baseline experiments. 
However, there are a large number of degeneracies among the cosmological parameters. Depending on the decay scenario taken into account (for example, whether neutrinos decay while they are relativistic or non-relativistic in the early universe, or if the interactions that trigger the decay are time dependent, etc.) the bounds maybe relaxed. A detailed discussion of such scenarios is beyond the scope of this article and from here-on we implicitly assume that terrestrial bounds are substantially meaningful.

In this paper we want to assess the capability of the DUNE experiment to constrain the quantity $\beta_3$ and check to which precision it could be measured in the chance of $\beta_3 < \infty$. Normal ordering of the neutrino states is assumed throughout the rest of the paper. Compared to other studies \cite{Choubey:2017dyu}, we improve the analysis in the following details:
\begin{itemize}
 \item we include the $\nu_\tau$ appearance channel \cite{deGouvea:2019ozk,Ghoshal:2019pab} in both leptonic and hadronic decay modes; this channel has been shown to be very promising in constraining parameters of new physics models;
 \item we include in the analysis the neutral current channel contributions \cite{Coloma:2017ptb} since in the ID scenario the number of active neutrinos is not conserved during propagation; 
 \item for the analytic understanding of our results,  we present (to our knowledge for the first time in the literature) the muon-flavored neutrino appearance and disappearance vacuum probabilities expanded up to second-order in the small quantity $\alpha = \Delta m_{21}^2 L/ 2 E$  in the presence of a decaying $\nu_{3}$ and quantify how much (percentage) the corrections contribute order by order.
\end{itemize}
We find that a setup contemplating $\nu_\tau$ and NC events can give the following 90\% CL bound:
\begin{equation}
\label{bounbsb3_nostro}
\beta_3 > 5.1 \times 10^{-11} \;{\rm s/eV\qquad  (this~~work)}\,.
\end{equation}
This limit is more than 15\% larger than the one obtained for DUNE with a longer exposure in \cite{Choubey:2017dyu}, $\beta_3>4.5 \times 10^{-11}~s/eV$, and is competitive with bounds obtainable by other future neutrino experiments probing different $L/E$~\cite{Tang:2018rer,deSalas:2018kri,Choubey:2017eyg,Abrahao:2015rba}. 

Besides the standard neutrino flux (with an energy peak around $E\sim 2.5$ GeV \cite{Acciarri:2016crz, Acciarri:2015uup}) we have also analyzed the limits 
on $\beta_3$ achievable with a {\it tau optimized flux} \cite{Bishai, fluxes}, especially dedicated to maximize the number of available taus in DUNE; the bound is roughly as small as half the limit in Eq.(\ref{bounbsb3_nostro}): 
 $\beta_3>2.8 \times 10^{-11}~s/eV$. This is a consequence of the fact that this flux has worse performances on the $\nu_e$ appearance and $\nu_\mu$ disappearance channels, which nullify the benefit of having a huge sample of $\tau$ neutrinos \cite{Ghoshal:2019pab}.

This paper is organized as follows: in the next section we review the theory of neutrino decay and discuss the leading order transition probabilities useful for our study; in Sect.\ref{energyspectra} we discuss the technical details of our numerical simulation of the DUNE experiment and report on the neutrino energy spectra from charged current (CC) and neutral current (NC) interactions at different $\beta_3$ values; in Sect.\ref{sensitivity} we present our results about the DUNE sensitivity to the decay parameters; eventually, Sect.\ref{concl} is devoted to our conclusions. In addition, three appendices are included, where we report the transition probabilities up to $\alpha^2$ (appendix A), the (perturbatively evaluated) charged current event rates as a function of $\beta_3$ (appendix B) and the charged current energy spectra with the related bound on $\beta_3$ for the $\tau$ optimized flux (appendix C).

\section{Transition Probabilities in the case of Neutrino Decay}

In the invisible decay framework, our working hypothesis is that the third neutrino mass eigenstate, the heaviest in the Normal Hierarchy case, can decay into a lighter sterile eigenstate $\nu_4$ ($\nu_3 \rightarrow \nu_4 + S$) and another undetected particle $S$.
Thus, the flavor and mass eigenstates are related through a $4 \times 4$ mixing matrix as:
\begin{equation}\label{mixing}
\begin{pmatrix}
\nu_\alpha\\
\nu_s
\end{pmatrix}
= 
\begin{pmatrix}
U_{PMNS} & 0 \\
0 & 1  
\end{pmatrix}
\begin{pmatrix}
\nu_i\\
\nu_4
\end{pmatrix}\,,
\end{equation}
where the $U_{PMNS}$ is the usual $3\times3$ neutrino mixing matrix among active states.
 
Even though the sterile eigenstate is not directly involved in the neutrino mixing, it is clear that the lifetime of the third eigenstate modify the Schroedinger time-evolution equation of the neutrino mass states. Taking into account also the standard matter effects, the Hamiltonian of the system is indeed altered in the following way \cite{Berryman:2014yoa}: 
\begin{equation}\label{evol}
H = U\left[\frac{1}{2E}
\begin{pmatrix}
0&0&0\\
0&\Delta m^2_{21}&0\\
0&0&\Delta m^2_{31}
\end{pmatrix}
-i\frac{1}{2 \beta_3 E }
\begin{pmatrix}
0&0&0\\
0&0&0\\
0&0&1
\end{pmatrix}\right] U^\dagger
+ 
\begin{pmatrix}
      A&0&0\\0&0&0\\0&0&0
     \end{pmatrix}
\,,
\end{equation}
where 
the term $A = 2 \sqrt{2} G_F n_e E$ is the neutrino electron scattering in matter,
$G_F$ is the Fermi constant, $E$ the energy of the neutrino, and $n_e$ the electron density. 

In a long-baseline experiment, the impact of the matter potential  on the oscillation probabilities depends on the $L/E$ ratio. In the case of a matter density of 4 $g/cm^3$, the comparison of the energy behavior of the oscillation probabilities with and without matter effects (in the standard case of stable neutrinos) for an approximately 1300 km baseline  shows the largest difference in the $\nu_e$ appearance probability, where the vacuum case can be roughly 9\% smaller for energies around 2 GeV \footnote{An effect of the same order of magnitude can be seen in the $\bar{\nu}_e$ appearance probability, where the matter potential decreases the transition amplitude.}. However, the related effect on the number of events is not so relevant since the integral of the probability between 0.2 and 15 GeV is only 5\% larger in the matter than in the vacuum case; this difference does not change much even when $\beta_3$ is finite \cite{Ascencio-Sosa:2018lbk}. Moreover, since the correlation between the decay parameter and the matter potential is negligible, all  modifications to the probability shape due to the matter effect are basically unaltered by the presence of a finite $\beta_3$. Finally, matter potential has only a negligible effect on the disappearance probability. Thus, we expect that in the invisible decay model, the DUNE sensitivity to the decay parameter would not be drastically influenced by matter effect. For this reason, at least in the following analytical description we will consider for simplicity the vacuum approximation only. 

We now derive the oscillation probabilities in the case of unstable $\nu_3$ eigenstate. If $\lambda_i$'s are the eigenvalues of the Hamiltonian matrix and S is the diagonalizing matrix, the transition amplitude can be obtained in the following way:
\begin{equation}
  \braket{\nu_\beta|\nu_\alpha}=\sum_i S_{\beta i}S^{-1}_{i \alpha} e^{-i \lambda_i L} \,,
\label{amplitude1}    
\end{equation}
where $L$ is the distance travelled by the neutrino after its creation. Notice that the hamiltonian $H$ in Eq.(\ref{evol}) is non-Hermitian, thus the inverse of S must appear in the amplitude.
The expression of the resulting transition probabilities is quite cumbersome, therefore we prefer to present the $\nu_\mu \to \nu_f$ oscillation formulae (with $f=e,\mu,\tau$) expanded  up to the second order in the parameter $\alpha=\frac{\Delta m^2_{21}}{2E}L$. We separate the various terms according to the convention $P_{\mu f}=P_{\mu f}^{(0)}+\alpha P_{\mu f}^{(1)}+\alpha^2 P_{\mu f}^{(2)}$, where the superscripts $(0), (1), (2)$ refer to the respective perturbative order. For the sake of simplicity, we quote here the zeroth-order results only, which capture the main effects of the decay, while deferring a discussion on the other terms (and a study of the goodness of our perturbative expansion) to Appendix A.
For the $\nu_\mu \to \nu_e$ transition we obtain:
\begin{equation} 
\begin{aligned}
P_{\mu e}^{(0)}= & \sin^2 {2\theta_{13}} \sin^2{\theta_{23}} \bigg [e^{-\frac{1}{\beta_3}\frac{L}{2E}} \,\sin^2\left(\frac{\Delta m^2_{31} L}{4E}\right)+ \bigg(\frac{1-e^{-\frac{1}{\beta_3}\frac{L}{2E}}}{2}\bigg)^2\bigg]\,,
\end{aligned}
\label{mue_zero}
\end{equation}
while for  the $\nu_\mu \to \nu_\tau$ appearance we get:
\begin{equation} 
\begin{aligned}
P_{\mu \tau}^{(0)}= & \cos^4 {\theta_{13}} \sin^2{2 \theta_{23}} \bigg [e^{-\frac{1}{\beta_3}\frac{L}{2E}} \,\sin^2\left(\frac{\Delta m^2_{31} L}{4E}\right)+ \bigg(\frac{1-e^{-\frac{1}{\beta_3}\frac{L}{2E}}}{2}\bigg)^2\bigg]\,.
\end{aligned}
\label{mutau_zero}
\end{equation}
Finally, for $\nu_\mu$ disappearance our result reads:
\begin{equation} 
\begin{aligned}
P_{\mu \mu}^{(0)}= &
 1+2 \left(e^{-\frac{1}{\beta_3}\frac{L}{2E}}-1\right) \cos^2{\theta_{13}} \sin^2 {\theta_{23}}+\left(e^{-\frac{1}{\beta_3}\frac{L}{2E}}-1\right)^2 \cos^4{\theta_{13}} \sin^4{\theta_{23}}\\ & -e^{-\frac{1}{\beta_3}\frac{L}{2E}}\left(\cos^4{\theta_{13}}\sin^2{2 \theta_{23}}+\sin^2{2 \theta_{13}} \sin^2{\theta_{23}}\right) \sin^2\left(\frac{\Delta m^2_{31} L}{4E}\right)\,.
\end{aligned}
\label{mumu_zero}
\end{equation}
We see that the decay parameter has two main roles. On the one hand, it acts as a damping factor, reducing the amplitudes by the quantity $e^{-\frac{1}{\beta_3}\frac{L}{2E}}$. On the other hand, it adds to the probabilities constant terms (i.e., terms which do not depend on the mixing angles) that contain the factor $\left( 1-e^{-\frac{1}{\beta_3}\frac{L}{2E}} \right)$. Thus, for small values of the decay parameters, we expect the appearance probabilities no longer to depend on the $L/E$ ratio and to converge to a fixed value (1/4 of the maximum value of the transition probability for both $\nu_\mu \to \nu_e$ and $\nu_\mu \to \nu_\tau$ transitions).
In the disappearance channel we observe again the same behaviour for small $\beta_3$ although the constant limiting value is approximated by $P_{\mu\mu}^{(0)}=(1-\cos^2\theta_{13}\sin^2\theta_{23})^2\sim 0.21$.
The explanation of this effect resides on the fact that, if the decay parameter is small, all neutrinos in the third mass eigenstate decay before reaching the far detector and since at leading order we are neglecting the mass difference between $\nu_1$ and $\nu_2$, the three neutrinos are no longer affected by oscillations.
We finally observe that, since the effect of decay is encoded in the damping factor which is common to every transition, all oscillation channels will be equally sensitive to the decay parameter. Thus a
collection of events in each channel can be very powerful in constraining $\beta_3$.

Notice that, in the presence of neutrino decay, Eqs. (\ref{mue_zero}), (\ref{mutau_zero}) and (\ref{mumu_zero}) imply:
\begin{equation}
    \sum_\alpha^{e,\mu,\tau} P_{\mu\alpha} =1+ (e^{-\frac{1}{\beta_3}\frac{L}{E}}-1) \cos^2{\theta_{13}}\sin^2{\theta_{23}}\ne 1\,;
\label{unitarity}
\end{equation}
indeed, if $\nu_3$ can decay into a sterile neutrino during its travel, the total number of active neutrinos will decrease when the distance travelled by the particles increases.
So we expect that the total number of active neutrinos will decay exponentially from the maximum, obtained when we are close to the neutrino source (small $L$), to an asymptotic value that depends at the leading order on $\theta_{23}$ and $\theta_{13}$ only.

\begin{figure}[t]
\begin{center}
\includegraphics[height=6.1cm,width=7.5cm]{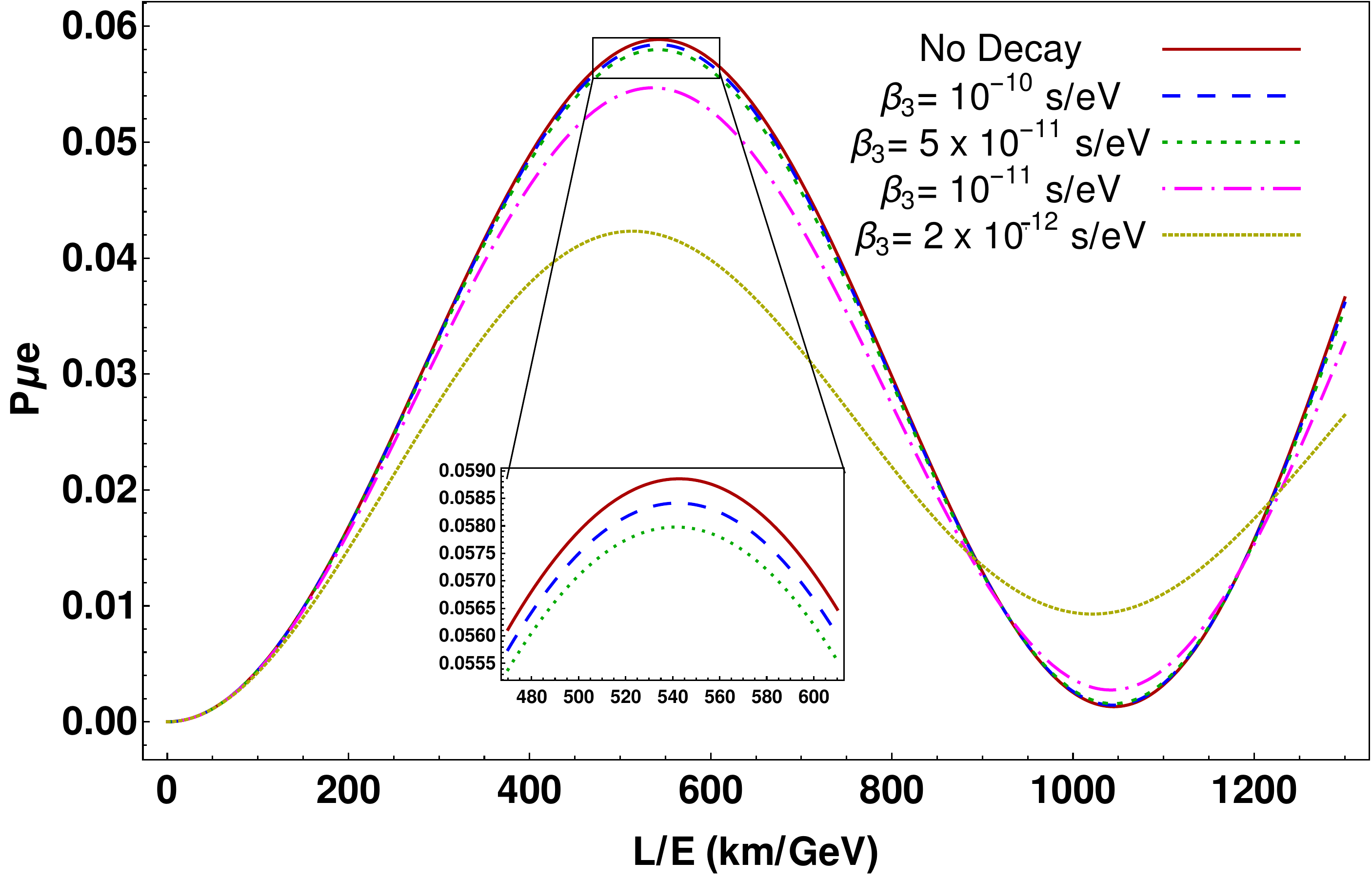} 
\includegraphics[height=6.1cm,width=7.5cm]{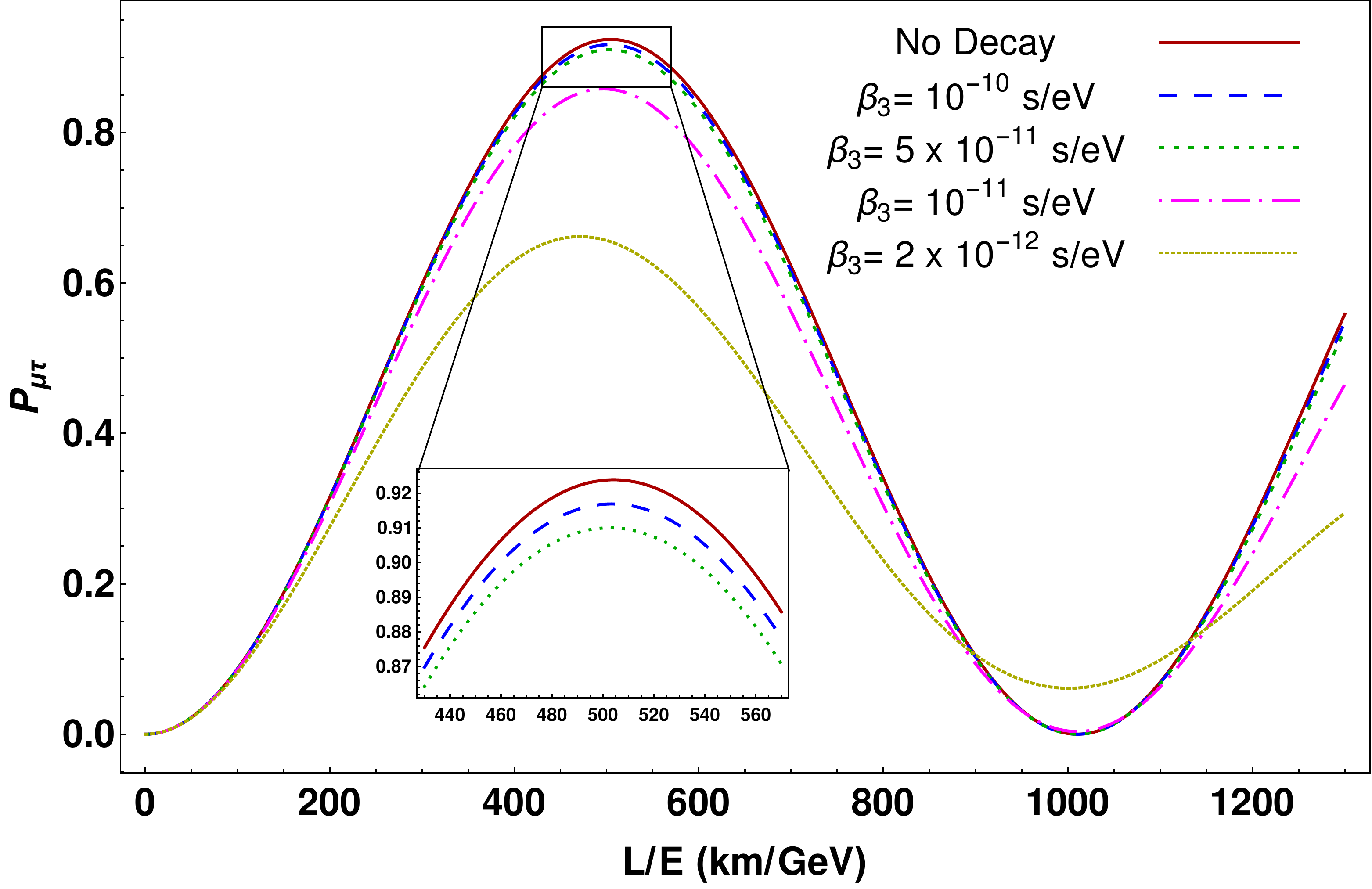} 
\includegraphics[height=6.1cm,width=7.5cm]{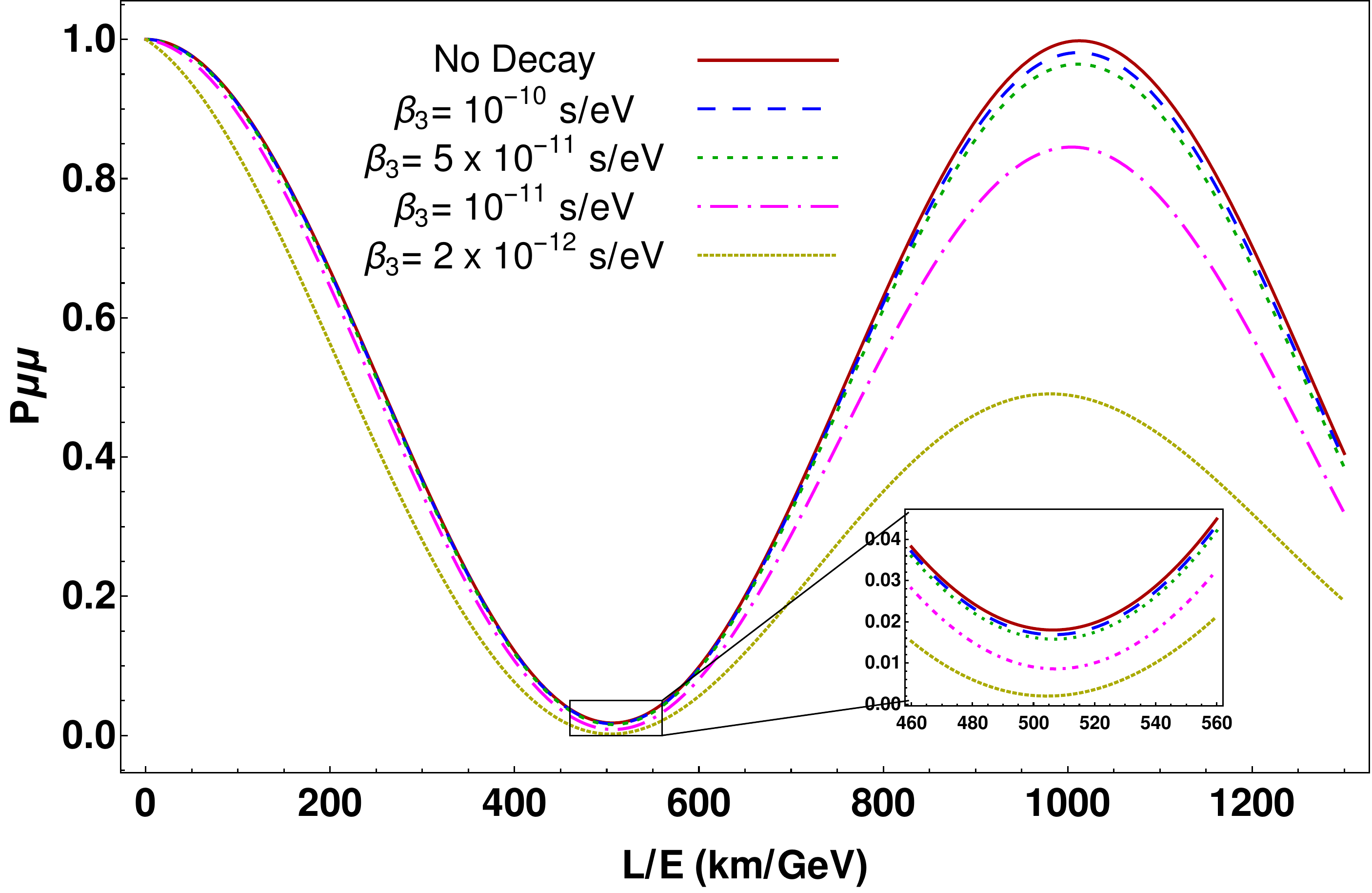} 
\caption{\it Exact $L/E$ dependence in vacuum of $P_{\mu e}$ (top-left panel), $P_{\mu \tau}$ (top-right panel) and $P_{\mu \mu}$ (bottom panel).  Different values of the decay parameter are shown: $\beta_3 = 10^{-10}~s/eV$ (blue dashed line), 
$5 \times 10^{-11}~s/eV$ (green dotted line), $ 10^{-11}~s/eV$  (magenta dot-dashed line) and  $2 \times 10^{-12}~s/eV$  (yellow densely dotted line). Red solid lines refers to the behavior of $P_{\mu f}$ in the absence of decay.} 
\label{probs}
\end{center}
\end{figure}

Plots showing the exact dependence in vacuum of $P_{\mu \alpha}$ on $L/E$ between 0 and 1300 km/GeV are reported in Fig.(\ref{probs}) for different values of the decay parameter: $\beta_3 = 10^{-10}~s/eV$ (blue dashed line), 
$5\times10^{-11}~s/eV$ (green dotted line), $ 10^{-11}~s/eV$ (magenta dot-dashed line) and  $2\times10^{-12}~s/eV$ (yellow densely dotted line). These values have been chosen to be of the order of the decay parameter limits set by oscillation experiments reported in Eqs.(\ref{bounbsb3}-\ref{bounbsb3_2}). For the sake of comparison, we also show with red solid lines the behavior in the absence of decay, that is in the standard three neutrino framework.
As it can be seen, the main effect of the decay parameter in the $L/E$ region accessible by long-baseline experiments like DUNE is a decrease of the probabilities around the atmospheric peak ($L/E \sim 500$ Km/GeV). This reduction is approximately 1.5\% when $\beta_3 = 10^{-10}~s/eV$ , 3\% when $\beta_3 = 5 \times 10^{-11}~s/eV$, 15\% when $\beta_3 = 10^{-11}~s/eV$ and 45\% when $\beta_3 = 2 \times 10^{-12}~s/eV$. 
The flattening of the probabilities previously discussed can be noticed around the valleys, where on the other hand $P_{\mu e}$ and $P_{\mu \tau}$ increase.

\section{The DUNE Experiment and the Neutrino Energy Spectra}
\label{energyspectra}
DUNE (Deep Underground Neutrino Experiment) will be a long-baseline neutrino experiment based in the USA \cite{Acciarri:2016crz, Acciarri:2015uup, Abi:2020wmh, Abi:2020evt}. The accelerator facility (that will provide a $\nu_\mu$ neutrino beam) will be built at Fermilab together with the Near Detector \cite{ND1, ND2, ND3}. Several different neutrino fluxes have been proposed, but the most studied one is peaked at a neutrino energy of $\sim$2.5 GeV. DUNE will also be able to run either in neutrino and antineutrino modes, probing oscillations of both particles and antiparticles. 
The far detector facility will be located at the SURF (Sanford Underground Research Facility), 1300 km away from the neutrino source. This detector will consist of four 10kt LAr-TPC modules. 

The expected performances of the far detector have been widely studied by the DUNE collaboration, which provided efficiency functions and smearing matrices for $\nu_e$ appearance and $\nu_\mu$ disappearance channels \footnote{Since the effect of the neutrino decay at very small $L/E$ is negligible, we did not include an explicit Near Detector in our analysis.}. A detailed description of the signal and the backgrounds for these two channels can be found in Refs. \cite{Acciarri:2016crz, Acciarri:2015uup}. We consider a DUNE running time of 3.5 years in neutrino mode and 3.5 years in antineutrino mode.

All the numerical simulations in this paper have been performed using the GLoBES software \cite{Huber:2004ka, Huber:2007ji} for which the DUNE collaboration provided ancillary files for the study of $\nu_e$ appearance and $\nu_\mu$ disappearance channels \cite{Alion:2016uaj}. The systematic uncertainties have been included as overall normalization errors, which are set to 2\% and 5\% for the $\nu_e$ and $\nu_\mu$ signals, respectively. All events (signal and backgrouds) are grouped into bins of 125 MeV size.
In addition, our numerical simulations will be supplemented by two more sources of events:
\begin{itemize}
\item the  $\nu_\tau$ appearance channel and the subsequent hadronic \cite{deGouvea:2019ozk} and electronic \cite{Ghoshal:2019pab} decay modes;
for the $\tau$ electronic decay we considered a 6\% overall detection efficiency for the signal, a signal-to-background ratio of 2.45, and a signal systematic uncertainty of 20\%, while for the $\tau$ hadronic decay we take into account that only 30\% of the hadronically decaying $\tau$-s are detected, with the 0.5\% of the NC events as a background;
\item  the neutral current channel, introduced in Ref. \cite{Coloma:2017ptb}. Accordingly, we have implemented the NC in GLoBES using an overall 90\% signal detection efficiency and a systematic uncertainty of 10\%; since the backgrounds come from the mis-identification of charged current events, we add to the background sample a conservative  10\% of the $\nu_\mu$ and $\nu_e$ CC events and all the $\nu_\tau$ CC events where the $\tau$ lepton decays hadronically. Considering that the number of active neutrinos is not conserved in the neutrino decay framework, we expect this channel to be very sensitive to the decay parameter (see Eq. (\ref{unitarity})).
\end{itemize}

\begin{table}[h!]
\centering
\begin{tabular}{|lcc|} \hline 
Parameter &    Central Value  & Relative Uncertainty  \\ \hline
$\theta_{12}$ & $33.82^\circ$ & 2.3\% \\ 
$\theta_{23}$ & $48.3^\circ$  & 2.2\% \\ 
$\theta_{13}$ & $8.61^\circ$  & 1.4\% \\
$\delta_{CP}$  & $222^\circ$ & 13\% \\
$\Delta m^2_{21}$ & 7.39$\times10^{-5}$~eV$^2$ & 2.8\% \\ 
$\Delta m^2_{31}$  & 2.523$\times10^{-3}$~eV$^2$ &  1.3\% \\ 
\hline 
\end{tabular}
\caption{\label{tab:oscpar_nufit2} \it Best fit value and relative uncertainty of neutrino oscillation parameters used in our simulation from a global fit to neutrino oscillation data \cite{Esteban:2018azc}.}
\end{table}

\subsection{Energy Spectra of Detected Neutrinos}

\begin{figure}[h!]
\begin{center}
\includegraphics[height=6.1cm,width=7.5cm]{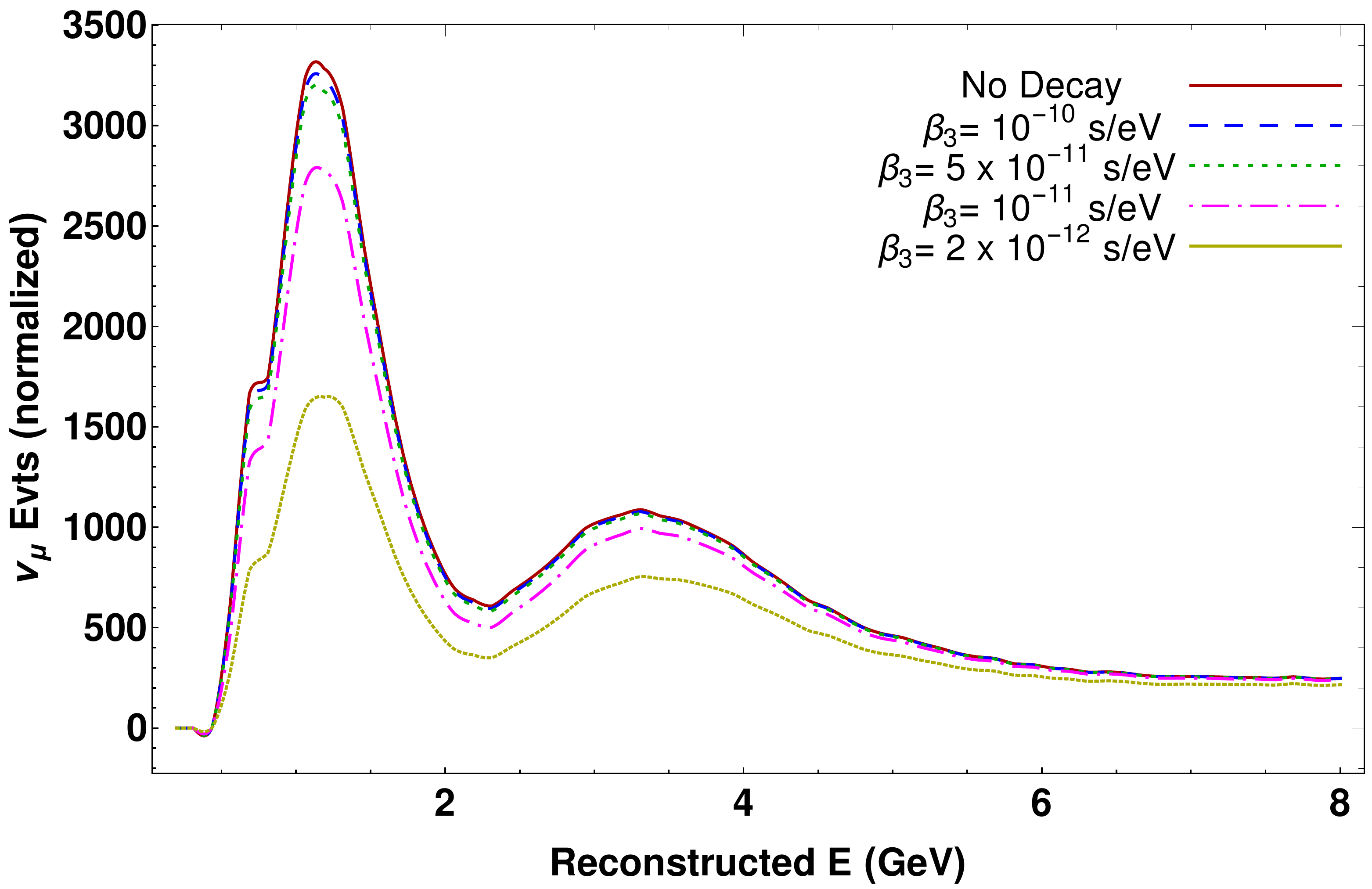} 
\includegraphics[height=6.1cm,width=7.5cm]{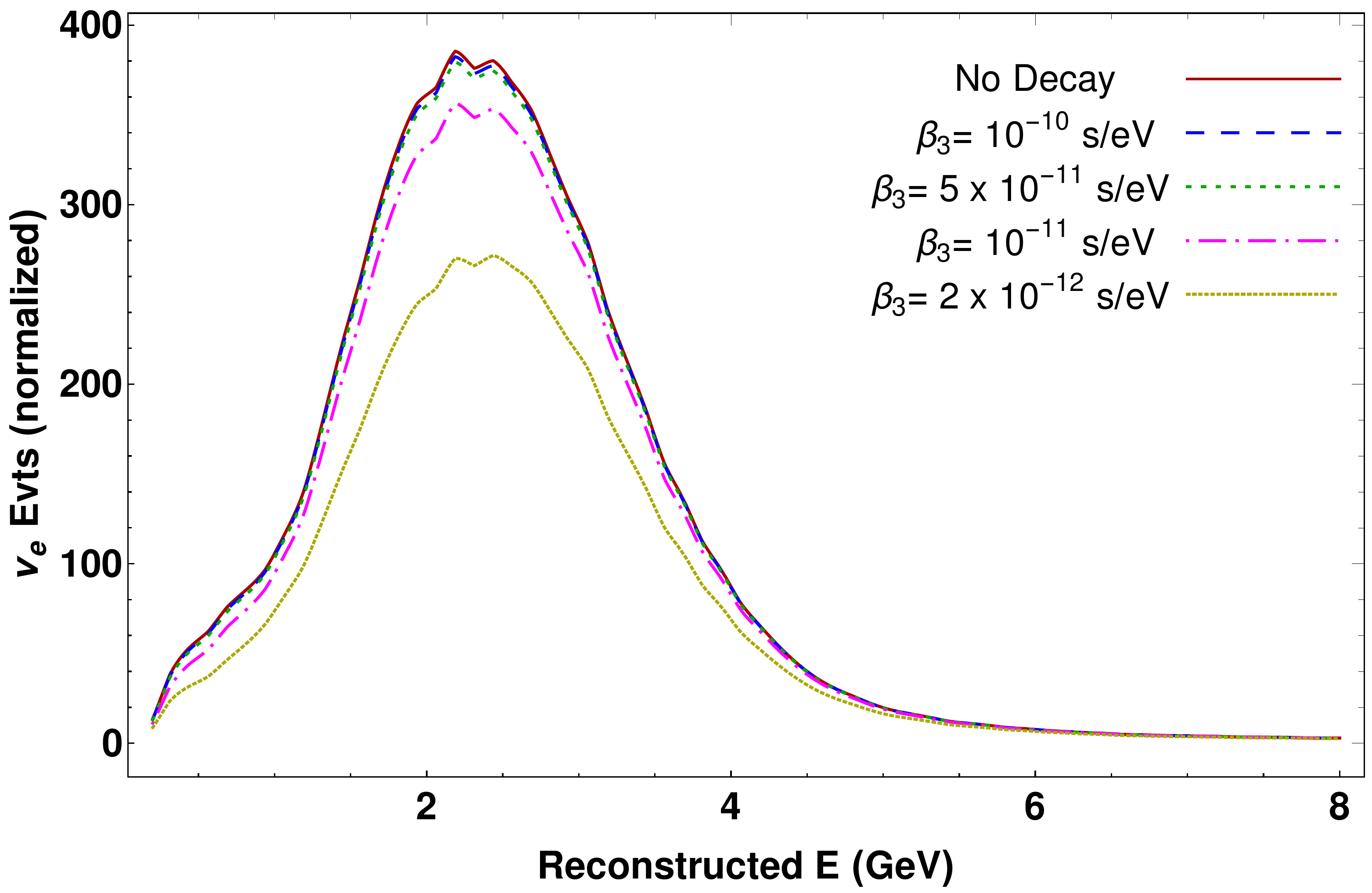} 
\includegraphics[height=6.1cm,width=7.5cm]{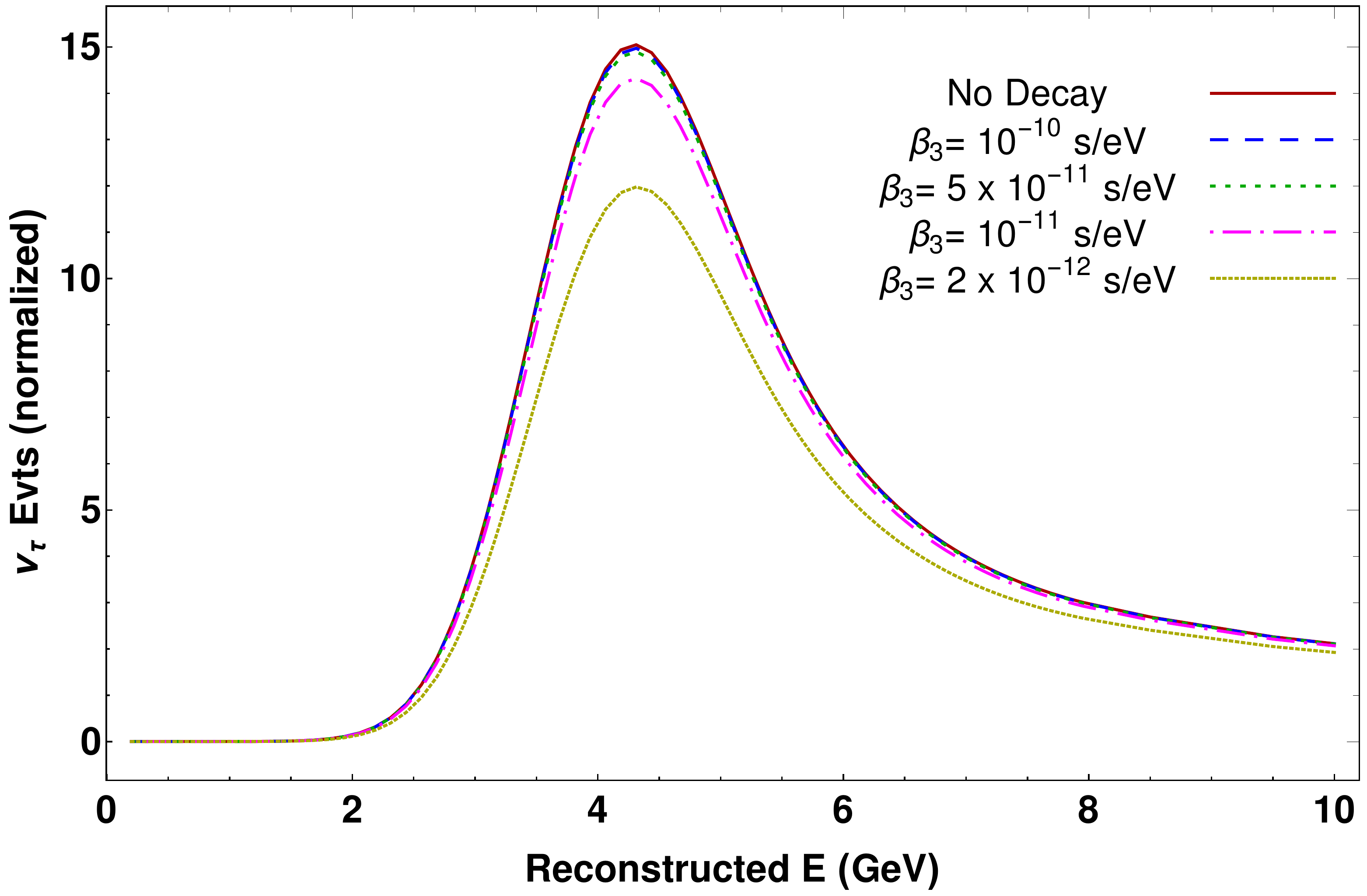} 
\includegraphics[height=6.1cm,width=7.5cm]{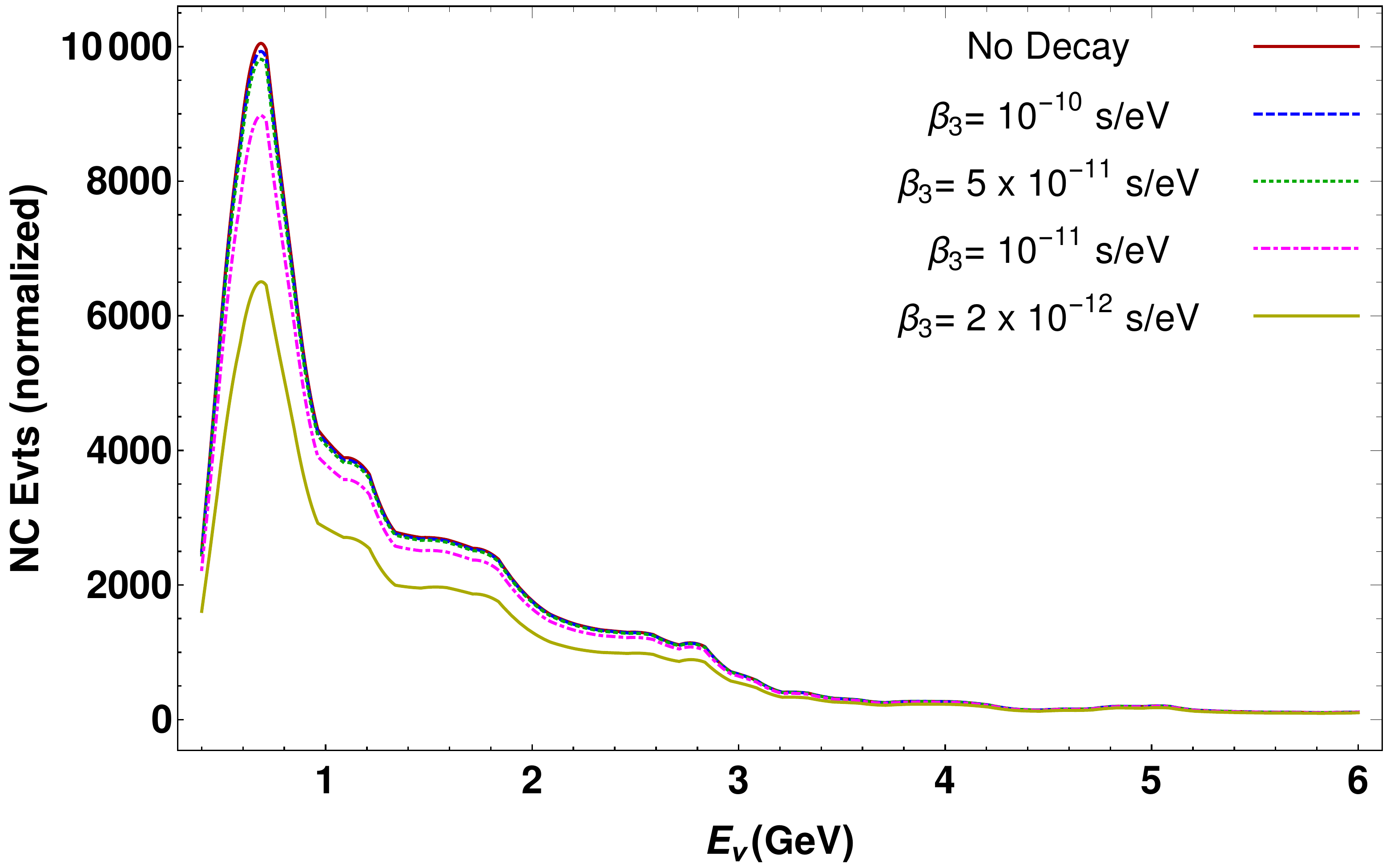} 
\caption{\it Expected $\nu_\mu$, $\nu_\tau$ and $\nu_e$ CC and NC events in DUNE as a  function of the reconstructed neutrino energy for different values of the decay parameter (same styles as in Fig.(\ref{probs})). The number of events on the y-axis has been normalized using the variable bin width given by the collaboration in the GLoBES configuration files \cite{Alion:2016uaj}. } 
\label{DUNE_std}
\end{center}
\end{figure}

With the final goal to show the DUNE sensitivity to the decay parameter, in this section we discuss how the energy spectra of the detected neutrinos in each oscillation channel are influenced by a finite $\beta_3$.
Rates are computed using the best fits for the oscillation parameters reported in Tab.(\ref{tab:oscpar_nufit2}) \footnote{Other equally valid global fits to neutrino oscillation data can be found in \cite{deSalas:2018bym}.}.

In Fig.(\ref{DUNE_std}) we show the number of $\nu_\mu$ (upper left panel), 
$\nu_e$ (upper right panel), $\nu_\tau$ (lower left panel) CC events  as well as NC events (lower right panel), as a function of the reconstructed neutrino energy for the four different values of the decay parameter $\beta_3 = (0.2, \, 1, \, 5, \, 10) \times 10^{-11}$~s/eV (non-continuous lines), and in the standard three neutrino framework (solid line). 

The effect of the decay parameter on the CC spectra is a decrease in the number of events for every value of the reconstructed neutrino energy, with a shape reproducing the behavior implied by the oscillation probabilities, as shown in Fig.(\ref{probs}). Thus, for example, a maximum in $P_{\mu e}$ around $L/E\sim 500$ Km/GeV translates into a peak in the number of $\nu_e$ CC events at  $E \sim 2.5$ GeV. A similar scenario is observed in the number of $\nu_\mu$ CC where the spectrum presents a valley around 2 GeV that corresponds to the minimum in the disappearance probability.

The NC spectrum shows the same dependence on $\beta_3$, but presents also a remarkable decrease in  the number of expected events at high energies. This is mainly due to the wrong reconstruction of the neutrino energy. Indeed, in the NC events the neutrino energy is often underestimated, as it can be deduced from the smearing matrices provided in \cite{Acciarri:2015uup}. However, since the NC smearing matrices provided by the DUNE collaboration were not meant to be used for the NC signal but only for the NC background, they were obtained from simulations  with small statistics. Thus, the main effect on our simulations of such matrices is to produce some unphysical structures in the spectra (as, for example, unwanted wiggles between 1 and 3 GeV). In order to avoid such a bad behavior, we smoothed out the smearing matrices and obtained more realistic NC spectra, compatible with the one obtained using the improved smearing matrices discussed in [57]. We want to underline that a more detailed study of the NC migration matrices should be carried out to properly take into account the NC events as signal events. However, the shapes of the spectra are not important in the determination of the decay parameter sensitivity because, as showed in Fig. (\ref{DUNE_std}), the effect of $\beta_3$ of the order of $10^{-11}~s/eV$ is a uniform decreasing of the number of events, with no spectral distortions. We repeated our analysis using different smearing matrices for the NC and $\nu_\mu$ CC channels and we  confirm this  property.

\section{DUNE Sensitivity to Neutrino Decay}
\label{sensitivity}
In this section we report on the ability of the DUNE experiment to set a lower bound on $\beta_3$ and on the precision $\beta_3$ can be measured assuming for it a finite value. 
All the following $\chi^2$ analyses are based on the pull-method described in Refs. \cite{Huber:2002mx,Fogli:2002pt,Ankowski:2016jdd,Meloni:2018xnk}.

\begin{figure}[t]
\begin{center}
\includegraphics[height=6.1cm,width=7.5cm]{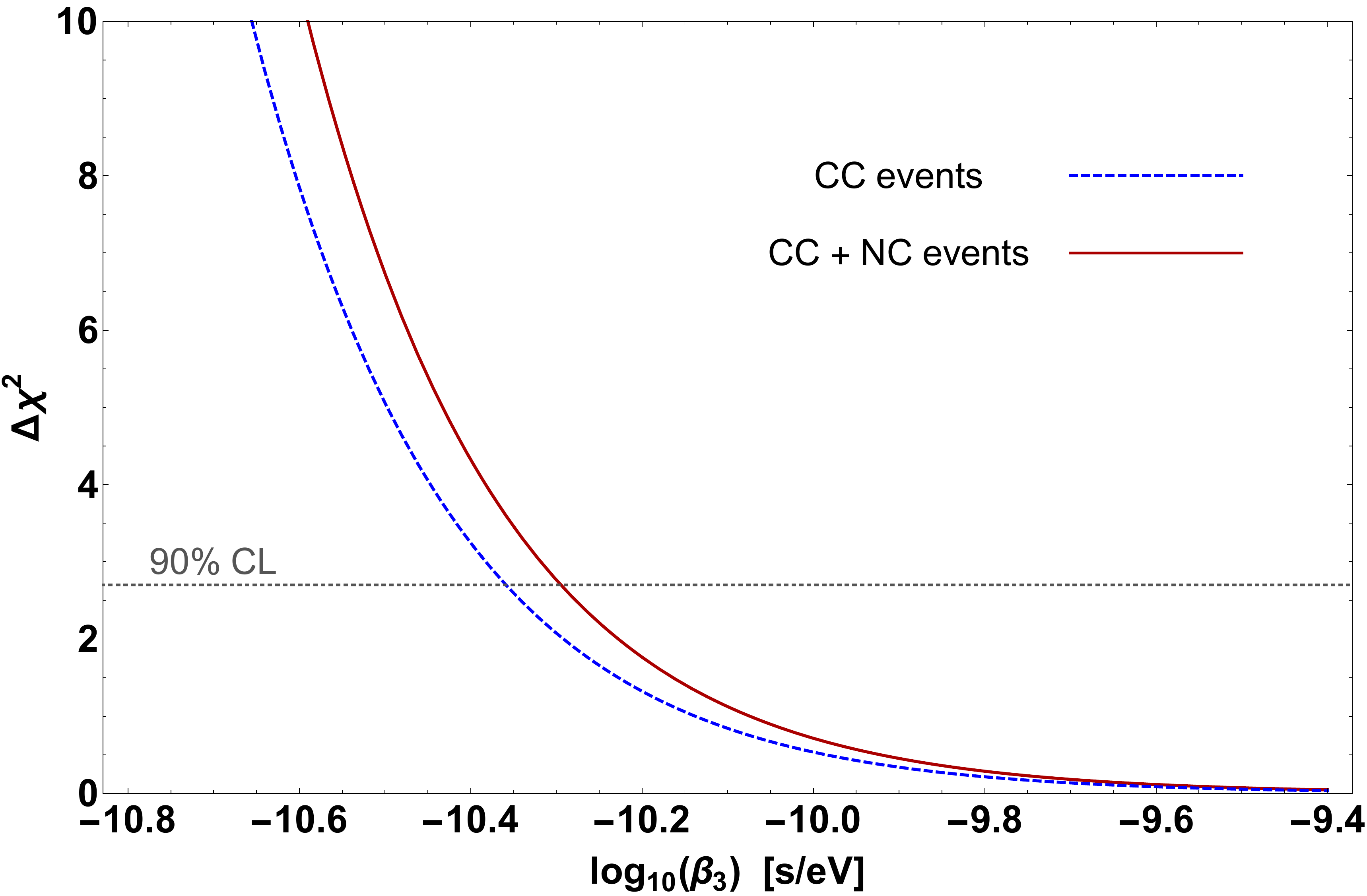} 
\caption{\it DUNE sensitivity to the decay parameter. The blue dashed line has been obtained using only the CC channels, while the red solid one has been obtained adding the NC channel. Here $\Delta \chi^2=\chi^2-\chi^2_{min}$.} 
\label{alpha_sens}
\end{center}
\end{figure}

\begin{figure}[t]
\begin{center}
\includegraphics[height=6.1cm,width=7.5cm]{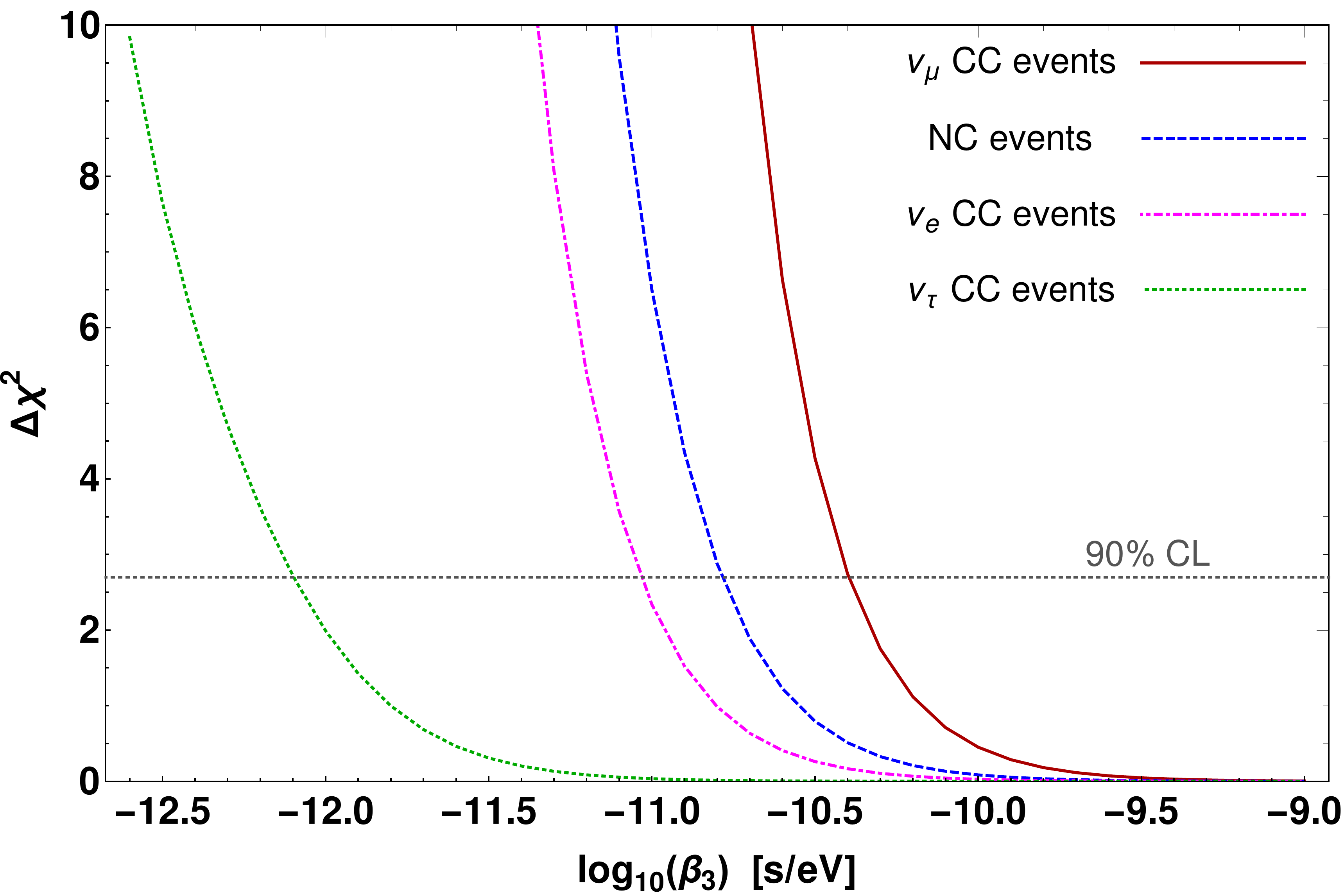} 
\caption{\it Contributions to $\beta_3$ by the different transition channels: red solid line refers to the $\nu_\mu$ CC events, the magenta dot-dashed line to the $\nu_e$ CC events, the green densely-dot-dashed line to the $\nu_e$ CC events while the blue dashed line represents the NC contribution. } 
\label{spacchettamento}
\end{center}
\end{figure}

In Fig.(\ref{alpha_sens}) we report our results for the sensitivity to $\beta_3$ when only CC (blue dashed line) and CC+NC events (red solid line) are taken into account. The curves have been obtained with true values of the standard oscillation parameters listed in Tab.(\ref{tab:oscpar_nufit2}). As fit values, we considered the same central points with their quoted uncertainties. For this analysis we used the full matter Hamiltonian showed in Eq. (\ref{evol}). \\
First of all, we notice that the addition of the NC events will be able to increase the lower bound on $\beta_3$ by roughly 16\%. In particular, the lower limit from the CC+NC analysis, $\beta_3 > 5.2 \times 10^{-11}$~s/eV, would be the best world limit set by a single long-baseline experiment. 
It is worth to mention that the limit set by the CC-only analysis (namely $\beta_3 > 4.4 \times 10^{-11}$~s/eV) is very similar to the one discussed in Ref. \cite{Choubey:2017dyu} where only $\nu_\mu$ disappearance and $\nu_e$ appearance channels (and a longer DUNE running time) were considered. This essentially means that the inclusion of the $\nu_\tau$ events in the analysis provides only a small contribution to the sensitivity, due to the (somehow) limited statistics. This fact can be appreciated in Fig.(\ref{spacchettamento}), where we split the contributions to the DUNE sensitivity to $\beta_3$ given by the different channels (see the caption for details). 
We see that the $\nu_\tau$ appearance is sensitive only to very small decay parameters, while the largest contribution comes from the $\nu_\mu \to \nu_\mu$ disappearance channel because, beside providing a larger number of interactions, the variation of the events as $\beta_3$ decrease  is larger than in the other channels (see Appendix B).

\begin{figure}[ht]
\begin{center}
\includegraphics[height=6.1cm,width=7.5cm]{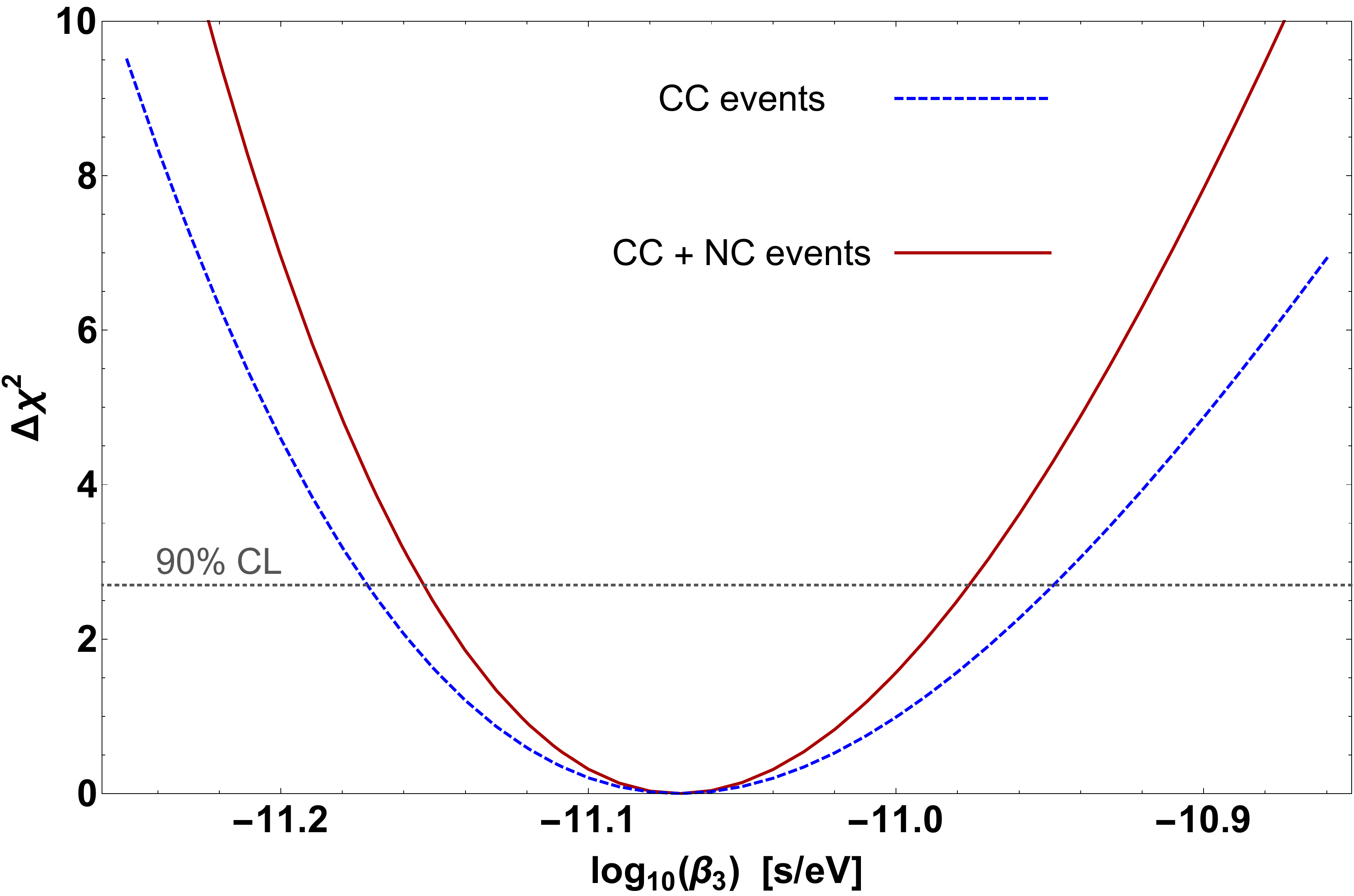}
\caption{\it $\Delta \chi^2$ as a function of $\beta_3$ obtained with a true value  $\beta_3=8.5\times 10^{-12}~s/eV$, corresponding to the best fit from 
MINOS and T2K data analysis \cite{Gomes:2014yua}. The blue dashed line has been obtained using only the CC channels, while the red solid  has been obtained adding the NC channel.} \label{alpha_sens2}
\end{center}
\end{figure}

As for a precision measurement of a possibly finite decay parameter, we show in Fig.(\ref{alpha_sens2}) an example in which the true value $\beta_3=8.5 \times 10^{-12}~s/eV$   (best fit obtained by MINOS and T2K data \cite{Gomes:2014yua}) is assumed; the numerical results highlight that roughly $23$\% and $20$\% precision can be achieved, if CC only or CC + NC are considered in the analysis, respectively.
Finally, in Tab.(\ref{tab1}) we collect both the 90\% CL bound and the 90\% CL error regions that DUNE will be able to set on $\beta_3$. For reference, we also included  the results for two more assumptions on the true $\beta_3$ value:  $\beta_3=1.2 \times 10^{-11}~s/eV$ (MINOS best fit) and $\beta_3=1.6 \times 10^{-12}~s/eV$ (T2K best fit \cite{Gomes:2014yua}).
We clearly see that the precision we can achieve varies from a maximum of $\sim 30$\% for $\beta_3 = 1.2 \times 10^{-11}~s/eV$ to a minimum of $\sim 10$\% for the smallest $\beta_3$; this is due to the fact that the difference among the values of a given transition probability computed at two different $\beta_3$'s is amplified in the case of small decay parameter (see Fig.(\ref{DUNE_std})).
\begin{table}
\centering
\begin{tabular}{c|c|c|}
\cline{2-3}
\multicolumn{1}{l|}{}                                    & \textit{\textbf{CC only}}                            & \textit{\textbf{CC+NC}}                              \\ \hline \hline
\multicolumn{1}{|c|}{\textit{$\beta_3=\infty$}}                  & $\beta_3$ \textgreater $4.4 \times 10^{-11}$ s/eV                        & $\beta_3$ \textgreater $5.1\times 10^{-11}$ s/eV                        \\ \hline
\multicolumn{1}{|c|}{$\beta_3=1.2 \times 10^{-11}$ s/eV}  & $ \beta_3 \in[0.91-1.78]\times 10^{-11}$ s/eV    & $\beta_3 \in [0.94-1.67]\times 10^{-11}$  s/eV  \\ \hline
\multicolumn{1}{|c|}{$\beta_3=8.5 \times 10^{-12}$ s/eV} & $\beta_3 \in[0.65-1.12]\times 10^{-11}$ s/eV  & $\beta_3 \in [0.70-1.05]\times 10^{-11}$ s/eV   \\ \hline
\multicolumn{1}{|c|}{$\beta_3=1.6 \times 10^{-12}$ s/eV} & $\beta_3 \in [1.46-1.83]\times 10^{-12}$ s/eV  & $\beta_3 \in[1.48-1.79]\times 10^{-12}$ s/eV  \\ \hline
\end{tabular}
\caption{\label{tab1} \it 90\% CL lower bound ($\beta_3=\infty$) and uncertainties on the decay parameter $\beta_3$ that will be set by DUNE when using the CC sample only (second column) or CC+NC events (last column).
Several assumptions on a finite $\beta_3$ are reported.
}
\end{table}

\section{Conclusions}
\label{concl}
DUNE will be one of the most important future neutrino oscillation experiments. It will collect a huge amount of events in every detection channel which allows not only to ameliorate the uncertainties on the standard mixing parameters (and possibly to determine the neutrino mass  hierarchy and the octant of $\theta_{23}$) but also to access to a whole series of phenomena not contemplated in the standard physics scenario.

In this paper we have described the DUNE capabilities in testing the invisible neutrino decay scenario, under the hypothesis that the mass eigenstates are normally ordered and the heaviest one $m_3$ is subject to decay to invisible particles including a sterile neutrino state. 
Although the $\nu_3$ lifetime has been already constrained in many ways using information from long and medium baseline as well as from atmospheric neutrino experiments, we showed that DUNE alone will be able to set the best  90\% CL long-baseline lower bound on the parameter $\beta_3=\tau_3/m_3$ ($\beta_3 > 5.1 \times 10^{-11}~s/eV$), performing an inclusive analysis where all charged current and neutral current channels are taken into account.
In the case $\beta_3$ would be measured before DUNE is operating, we have shown that an uncertainty of about $[10-30]$ \% can be set at 90\% CL, depending on the central value used.

As for the effects that a possible neutrino decay can have on the measurements of the leptonic CP phase $\delta_{CP}$ and the octant of the atmospheric angle, we have verified that $\beta_{3}$ above the  limits in Eqs.(\ref{bounbsb3}-\ref{bounbsb3_2}) will have a very marginal impact (and, for this reason, we refrained from presenting the corresponding plots). Also, in agreement with our analytical treatment of the probabilities, we verified that  $\beta_{3}$ has negligible correlations with the standard mixing parameters so that degeneracy regions in the standard parameter space are characterized by very large $\chi^2$ minima.

We conclude with the remark that, as it happened for MINOS, K2K and SuperKamiokande \cite{GonzalezGarcia:2008ru}, it is possible that a combined analysis of DUNE data with atmospheric and/or medium baseline experiments data could improve our limit in a significant way.

\section*{Acknowledgement}

A.G. thanks Amir Khan and Sudip Jana for helpful discussions. A.G. was supported by the research grant ``Ruolo del neutrino Tau in modelli di nuova Fisica nelle oscillazioni'' during the course of the project.

\newpage 

\section*{Appendix A: Transition probabilities up to second order in $\alpha=\frac{\Delta m^2_{21}}{2E}L$ in vacuum}
\label{appendixA}

\begin{figure}[h!]
\begin{center}
\includegraphics[height=6.1cm,width=7.5cm]{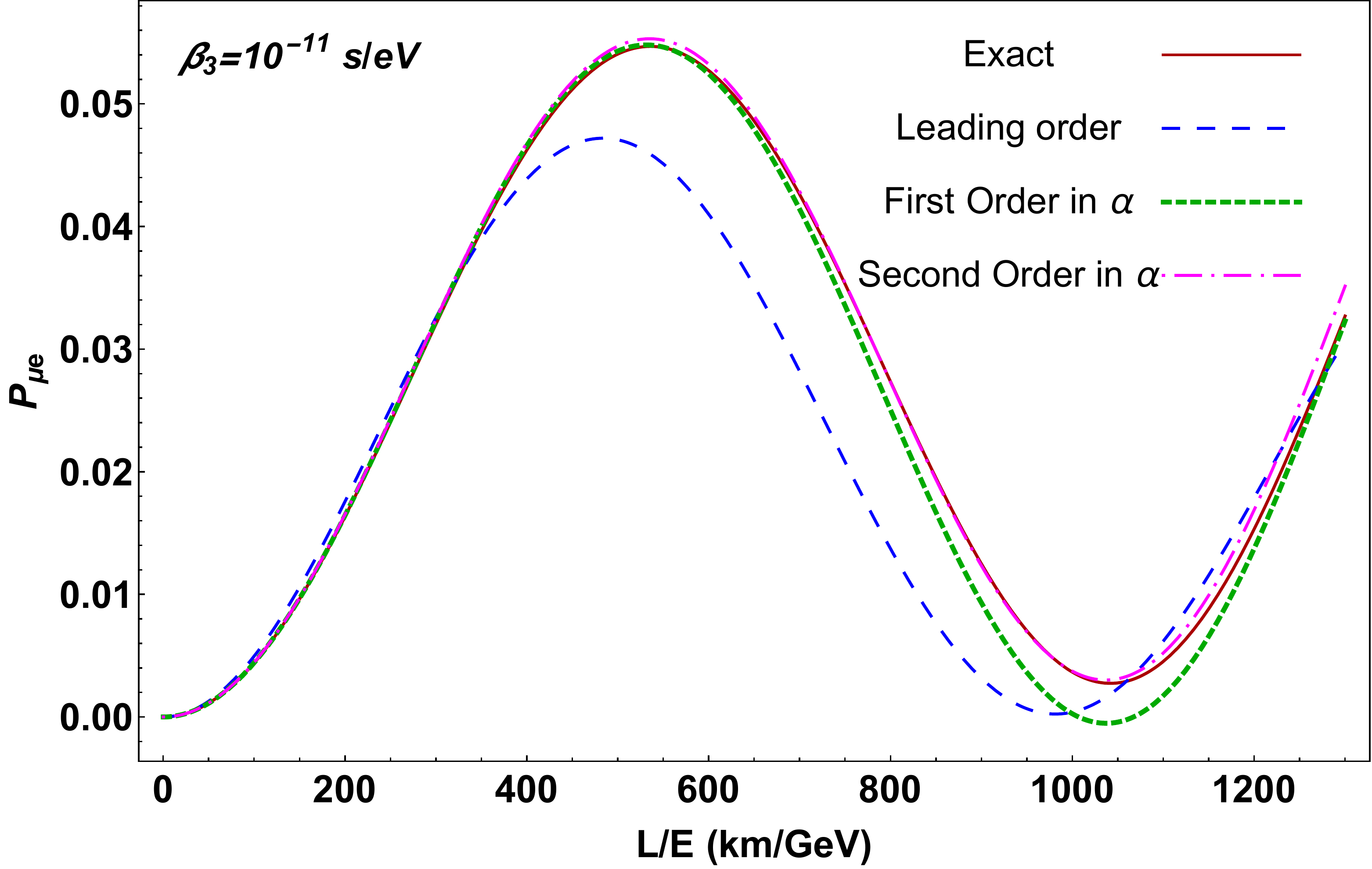} 
\includegraphics[height=6.1cm,width=7.5cm]{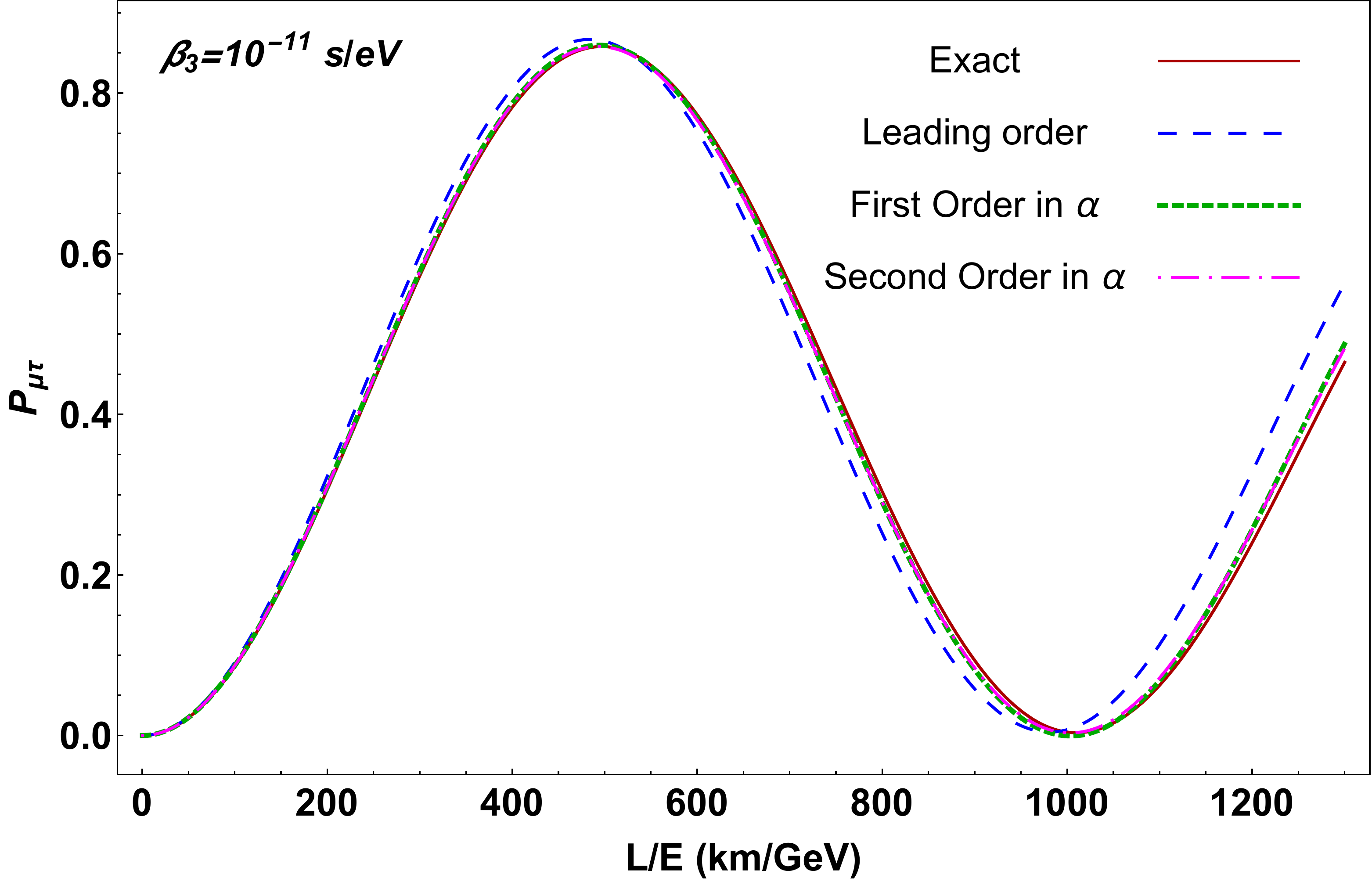} 
\includegraphics[height=6.1cm,width=7.5cm]{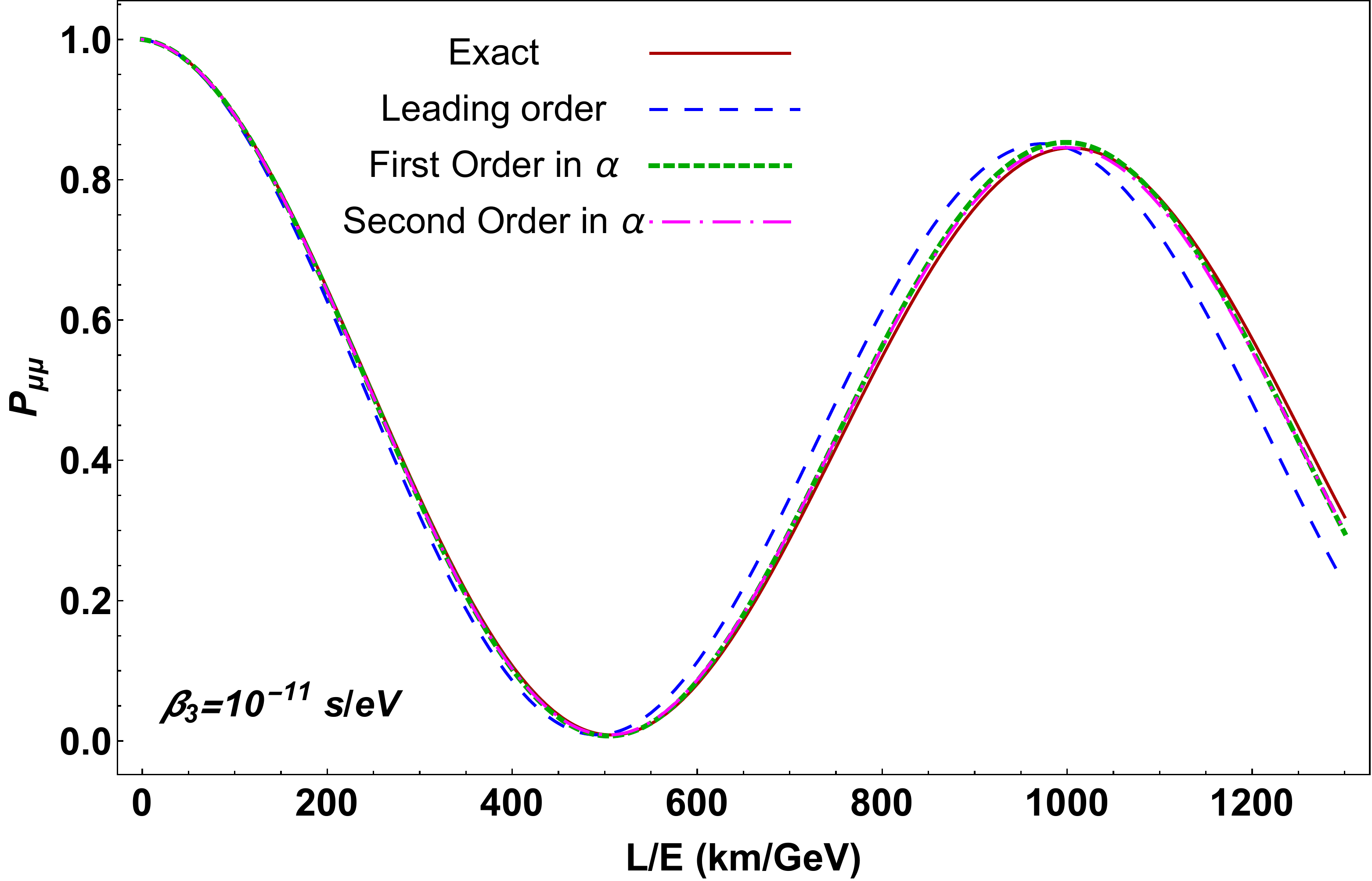} 
\caption{\it Comparison between vacuum exact $\nu_\mu \to \nu_\alpha$ oscillation probabilities (solid red line) and their expansions in $\alpha$ (blue dashed line for the leading order, green dotted line for the first order, magenta dot-dashed line for the second order) in the case of $\beta_3=10^{-11}$ $s/eV$. } 
\label{prob_confronto}
\end{center}
\end{figure} 
In this section we report the transition probabilities for $\nu_\mu \to \nu_{e,\mu,\tau}$ in the case of $\nu_3$ decay up to second order in $\alpha=\frac{\Delta m^2_{21}}{2E}L$ in the vacuum.  We assume an expansion form of the type  $P_{\mu f}=P_{\mu f}^{(0)}+\alpha P_{\mu f}^{(1)}+\alpha^2 P_{\mu f}^{(2)}$, where the superscript $(i)$ indicates the perturbative order taken into account.
For the $\nu_\mu \to \nu_e$ appearance we get:
\begin{eqnarray} 
&& \begin{aligned}
 P_{\mu e}^{(0)}= & \sin^2 {2\theta_{13}} \sin^2{\theta_{23}} \bigg [e^{-\frac{1}{\beta_3}\frac{L}{2E}} \,\sin^2\left(\frac{\Delta m^2_{31} L}{4E}\right)+ \bigg(\frac{1-e^{-\frac{1}{\beta_3}\frac{L}{2E}}}{2}\bigg)^2\bigg]\,,
\end{aligned}
\label{mue_zero_app}
\\
&& \begin{aligned}
P_{\mu e}^{(1)}=
 - & \, e^{-\frac{1}{\beta_3}\frac{L}{2E}} \cos^2{\theta_{13}} \sin{\theta_{23}} \sin{\theta_{13}} \bigg [ 2 \sin{\theta_{23}} \sin{\theta_{13}} \sin^2{\theta_{12}} \sin\left(\frac{\Delta m^2_{31} L}{2E}\right)\\
  &- \cos{\theta_{23}} \cos{\delta} \cos{2 \theta_{12}}\sin\left(\frac{\Delta m^2_{31} L}{2E}\right)\\ &+ \left ( e^{\frac{1}{\beta_3}\frac{L}{2E}}-1 \right) 
\cos{\theta_{23}} \sin{2\theta_{12}} \sin{\delta} +2\cos{\theta_{23}} \sin{\delta} \sin{2 \theta_{12}}\sin^2\left(\frac{\Delta m^2_{31} L}{4E}\right)  \bigg ]\,,
\end{aligned}
\label{mue_1}
\\
&&\begin{aligned}
P_{\mu e}^{(2)}=
  & \, \frac{e^{-\frac{1}{\beta_3}\frac{L}{2E}}}{8}  \cos^2{\theta_{13}} \sin{\theta_{12}} \bigg [ 8 \sin^2{\theta_{23}} \sin^2{\theta_{13}} \sin{\theta_{12}} \cos\left(\frac{\Delta m^2_{31} L}{2E}\right)\\
&- 4\sin{2 \theta_{23}} \sin{\theta_{13}} \cos{\theta_{12}} \cos\left(\frac{\Delta m^2_{31} L}{2E}+\delta \right)\\ 
&+ e^{\frac{1}{\beta_3}\frac{L}{2E}} \cos{\theta_{12}} \big (
4 \cos{2\theta_{12}} \sin{\theta_{13}} \sin{2 \theta_{23}} \cos{\delta} +  \sin{2 \theta_{12}} \\ 
& +3   \sin{2 \theta_{12}} \cos{2\theta_{23}}+2  \sin{2 \theta_{12}}  \cos{2 \theta_{13}} \sin^2{\theta_{23}} \big) \bigg ]\,,
\end{aligned}
\label{mue_second}
\end{eqnarray}

while for $\nu_\mu \to \nu_\tau$ appearance we obtain:
\begin{eqnarray} 
&&\begin{aligned}
P_{\mu \tau}^{(0)}= & \cos^4 {\theta_{13}} \sin^2{2 \theta_{23}} \bigg [e^{-\frac{1}{\beta_3}\frac{L}{2E}} \,\sin^2\left(\frac{\Delta m^2_{31} L}{4E}\right)+ \bigg(\frac{1-e^{-\frac{1}{\beta_3}\frac{L}{2E}}}{2}\bigg)^2\bigg]\,,
\end{aligned}
\label{mutau_zero_app}
\\
&&\begin{aligned}
P_{\mu \tau}^{(1)}=
 & \,\frac{ e^{-\frac{1}{\beta_3}\frac{L}{2E}}}{16} \cos^2{\theta_{13}} \bigg \{  \bigg[ (2\cos {2 \theta_{13}}-6) \cos{2 \theta_{12}}-4 \cos^2{\theta_{13}}\bigg] \sin^2{2 \theta_{23}} \sin\left(\frac{\Delta m^2_{31} L}{2E}\right) \\ 
 & -4\cos{\delta} \sin{4 \theta_{23}} \sin{\theta_{13}}\sin{2\theta_{12}} \sin\left(\frac{\Delta m^2_{31} L}{2E}\right) \\ &
 +8\bigg[e^{\frac{1}{\beta_3}\frac{L}{2E}}-\cos\left(\frac{\Delta m^2_{31} L}{2E}\right) \bigg] \sin{2 \theta_{23}} \sin{\theta_{13}} \sin{2 \theta_{12}} \sin{\delta} \bigg \}\,,
\end{aligned}
\label{mutau_first}
\\
&&\begin{aligned}
P_{\mu \tau}^{(2)}=
  & \, \frac{1}{256} \bigg \{ 8 \sin{\theta_{13}} \bigg[ \cos^2{\theta_{12}} \big(3+5\cos{2\theta_{12}}\big)\sin^2{2\theta_{23}}\sin{\theta_{13}} \\ &
 +16 \cos^3{\theta_{12}}\cos{\delta}\sin{4\theta_{23}}\sin{\theta_{12}}-16\cos{\theta_{12}}\cos{\delta}\sin{4\theta_{23}}\sin^2{\theta_{13}}\sin^3{\theta_{12}} \\ &
 +8 \sin^2{2\theta_{23}}\sin{\theta_{13}}\sin^4{\theta_{12}} \bigg]  
 -32\cos^2{\theta_{13}} \cos{\delta}\sin{4\theta_{23}}\sin{\theta_{13}}\sin{2\theta_{12}}\\ &-\bigg[ 21\cos{2\theta_{13}}-14+\cos{4\theta_{23}} \big ( 11\cos{2\theta_{13}}-18 \big) \\ & +2\sin^2{2\theta_{23}}\big(\cos{4\theta_{23}}+16 \cos{2\delta} \sin^2{\theta_{13}} \big) \bigg] \sin^2{2 \theta_{12}} \\ &
 -32 e^{-\frac{1}{\beta_3}\frac{L}{2E}} \cos{\theta_{13}} \bigg[ \big(8\cos^2{\theta_{23}} \cos^2{\theta_{12}}\sin^2{\theta_{23}}-2\sin{\theta_{13}}\sin^2{2 \theta_{23}}\sin{\theta_{13}}\sin^2{\theta_{12}} \\ &+\sin{\theta_{23}}\cos{\delta}\sin{4\theta_{23}}\sin{2\theta_{12}} \big)\cos\left(\frac{\Delta m^2_{31} L}{2E}\right) \\& +2\sin{2\theta_{23}}\sin{\theta_{13}}\sin{2\theta_{12}}\sin{\delta}\sin\left(\frac{\Delta m^2_{31} L}{2E}\right) \bigg] \bigg\}\,.
\end{aligned}
\label{mutau_second}
\end{eqnarray}

Finally the  $\nu_\mu \to \nu_\mu$ disappearance reads:
\begin{eqnarray} 
&&\begin{aligned}
P_{\mu \mu}^{(0)}= &
 1+2 (e^{-\frac{1}{\beta_3}\frac{L}{2E}}-1) \cos^2{\theta_{13}} \sin^2 {\theta_{23}}+(e^{-\frac{1}{\beta_3}\frac{L}{2E}}-1)^2 \cos^4{\theta_{13}} \sin^4{\theta_{23}}\\ & -e^{-\frac{1}{\beta_3}\frac{L}{2E}}(\cos^4{\theta_{13}}\sin^2{2 \theta_{23}}+\sin^2{2 \theta_{13}} \sin^2{\theta_{23}}) \sin^2\left(\frac{\Delta m^2_{31} L}{4E}\right)\,,
\end{aligned}
\label{mumu_zero_app}
\\
&&\begin{aligned}
P_{\mu \mu}^{(1)}=  & e^{-\frac{1}{\beta_3}\frac{L}{2E}} \cos^2{\theta_{13}} \sin^2{\theta_{23}} \big( 2 \cos^2{\theta_{23}} \cos^2{\theta_{12}}
+2 \sin^2 {\theta_{13}} \sin^2{\theta_{23}} \sin^2{\theta_{12}} \\ & - \cos{\delta} \sin{\theta_{13}}\sin{ 2\theta_{12}}\sin{2 \theta_{23}} \big)  \sin \left( \frac{\Delta m_{31}^2 L}{4E} \right)	\,,
\end{aligned}
\label{mumu_first}
\\
&&\begin{aligned}
P_{\mu \mu}^{(2)}=  & \frac{1}{4} \, \bigg\{ \big( 1+2\cos{2\theta_{12}} \big) \sin^2{2\theta_{23}}\sin^2{\theta_{13}}\sin^2{\theta_{12}}+ \\&
4\cos^2{\theta_{23}}\cos^2{\theta_{12}}\sin^2{\theta_{23}}\sin^2{\theta_{13}} \big( 2\cos{2\delta} \sin^2{\theta_{12}}-1 \big) \\&
- 4 e^{-\frac{1}{\beta_3}\frac{L}{2E}} \cos^2{\theta_{13}}\sin^2{\theta_{23}} \big( \cos^2{\theta_{23}}\cos^2{\theta_{12}} \\ &
-\cos{\theta_{23}}\cos{\delta}\sin{\theta_{23}}\sin{\theta_{13}}\sin{2\theta_{12}}+\sin^2{\theta_{23}}\sin^2{\theta_{13}}\sin^2{\theta_{12}} \big) \cos \left( \frac{\Delta m_{31}^2 L}{4E} \right)	\\ &
-\cos^4{\theta_{23}}\sin^2{2\theta_{12}}-\sin^4{\theta_{23}}\sin^4{\theta_{13}}\sin^2{2\theta_{12}} \\ &
-2\cos^3{\theta_{23}}\cos{\delta}\sin{\theta_{23}}\sin{\theta_{13}}\sin{4\theta_{12}}+\sin{2\theta_{23}} \cos{\delta} \sin^2{\theta_{23}}\sin{4\theta_{23}}\sin^3{\theta_{13}} \bigg\}\,.
\end{aligned}
\label{mumu_second}
\end{eqnarray}

In Fig.(\ref{prob_confronto}) we show the comparison between the exact vacuum probabilities (red solid lines) and the expansions discussed in this appendix for $\beta_3=10^{-11}$ $s/eV$ (blue dashed lines for the leading order, green dotted lines for the first order and magenta dot-dashed lines for the second order expansions). 
It is clear that around the first oscillation peak, which is the most important $L/E$ region for long baseline experiments, even the leading order is very accurate for the $\nu_\mu \to \nu_\mu$ and $\nu_\mu \to \nu_\tau$ probabilities. Indeed, the maximum discrepancy is roughly 2\% that can be further reduced by considering ${\cal O}(\alpha)$ and ${\cal O}(\alpha^2)$ terms. 

For the $\nu_\mu \to \nu_e$ probability the leading order expansion does not provide an accurate approximation. However, the inclusion of ${\cal O}(\alpha^2)$ contributions sensibly ameliorate the agreement with the exact result, with a maximum discrepancy of approximately 0.5\%. 
  
\section*{Appendix B: Charged and neutral current event rates}
\begin{figure}[h!]
\begin{center}
\includegraphics[height=6.1cm,width=7.5cm]{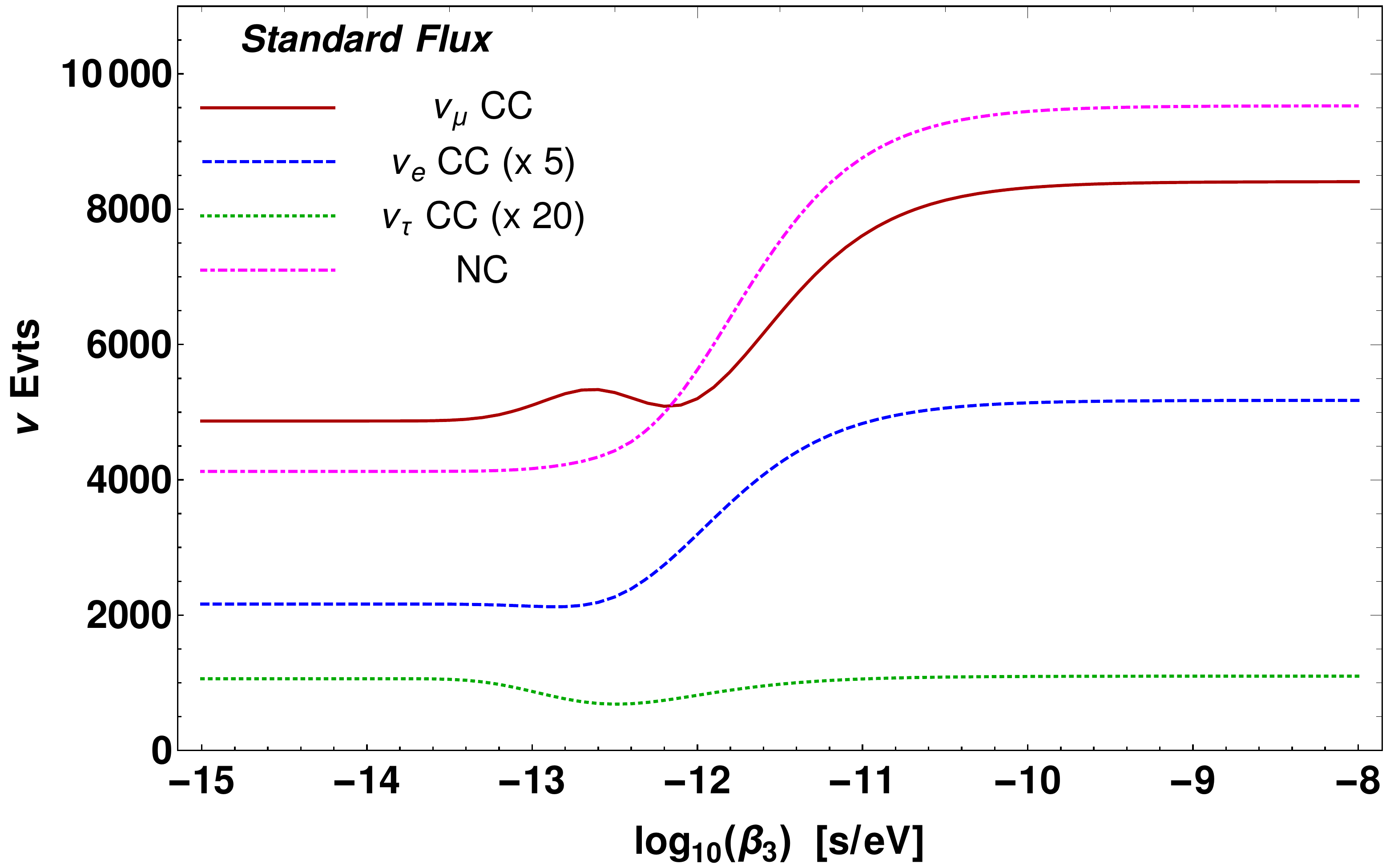} 
\includegraphics[height=6.1cm,width=7.5cm]{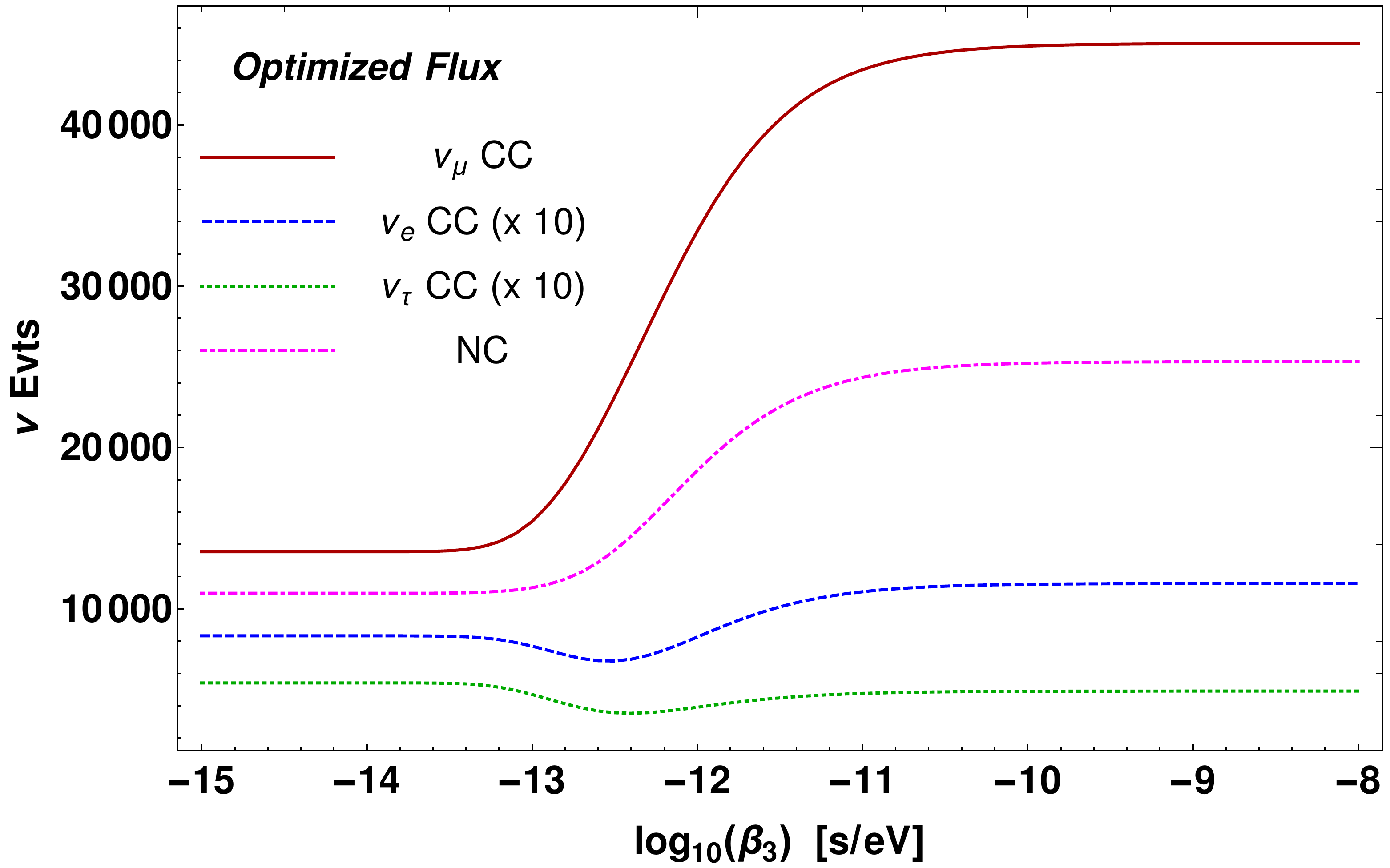} 
\caption{\it Expected $\nu_\mu$ CC (red solid lines), $\nu_e$ CC (blue dashed lines), $\nu_\tau$ CC (green dotted lines) and NC (magenta dot-dashed lines) events in DUNE with the standard (left panel) and the optimized (right panel) fluxes as a function of the decay parameter. For a better representation, the number of $\nu_e$ and $\nu_\tau$ events have been multiplied by a constant factor specified in the legend.} 
\label{DUNE_alpha_comp}
\end{center}
\end{figure}

For the sake of illustration, we report in Fig.(\ref{DUNE_alpha_comp}) the neutrino number of events in DUNE a function of $\beta_3$, under the assumption of standard and $\tau$ optimized neutrino fluxes (see the caption for details).

We can clearly see that the biggest change in the number of events can be found when $\beta_3\in[10^{-13}-10^{-11}]~s/eV$. In this region, for the standard flux, the number of $\nu_\mu$ and  $\nu_e$ CC events and NC  events decrease considerably reaching a constant plateau at smaller values of the decay parameter. On the other hand, in the case of the optimized flux, the number of $\nu_\tau$ CC and $\nu_e$ CC interactions only present a minimum around $\beta_3\sim\ 10^{-12}~s/eV$. 

This behaviour explains why other less performing long baseline experiments with neutrino energies of ${\cal O}$(GeV) have been able to set limits on $\beta_3$ of the order of $10^{-11}-10^{-12}~s/eV$. 
Notice also that the most relevant decrease in the number of events, for both fluxes, is seen in the $\nu_\mu$ CC and NC events which are then expected to contribute the most in constraining $\beta_3$.

\section*{Appendix C: Charged and neutral current energy spectra with the $\tau$ optimized flux}
\begin{figure}[ht]
\begin{center}
\includegraphics[height=6.1cm,width=7.5cm]{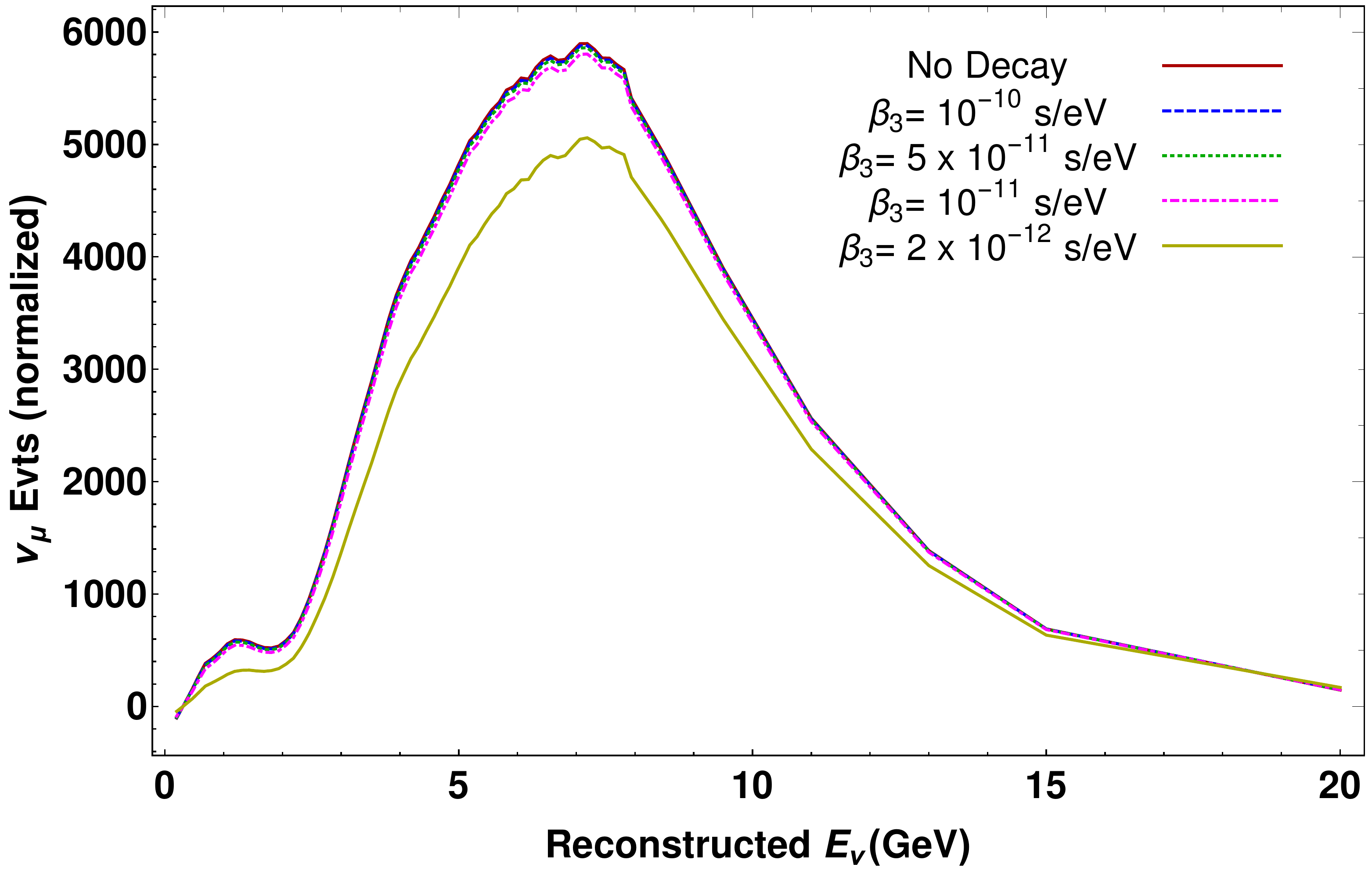} 
\includegraphics[height=6.1cm,width=7.5cm]{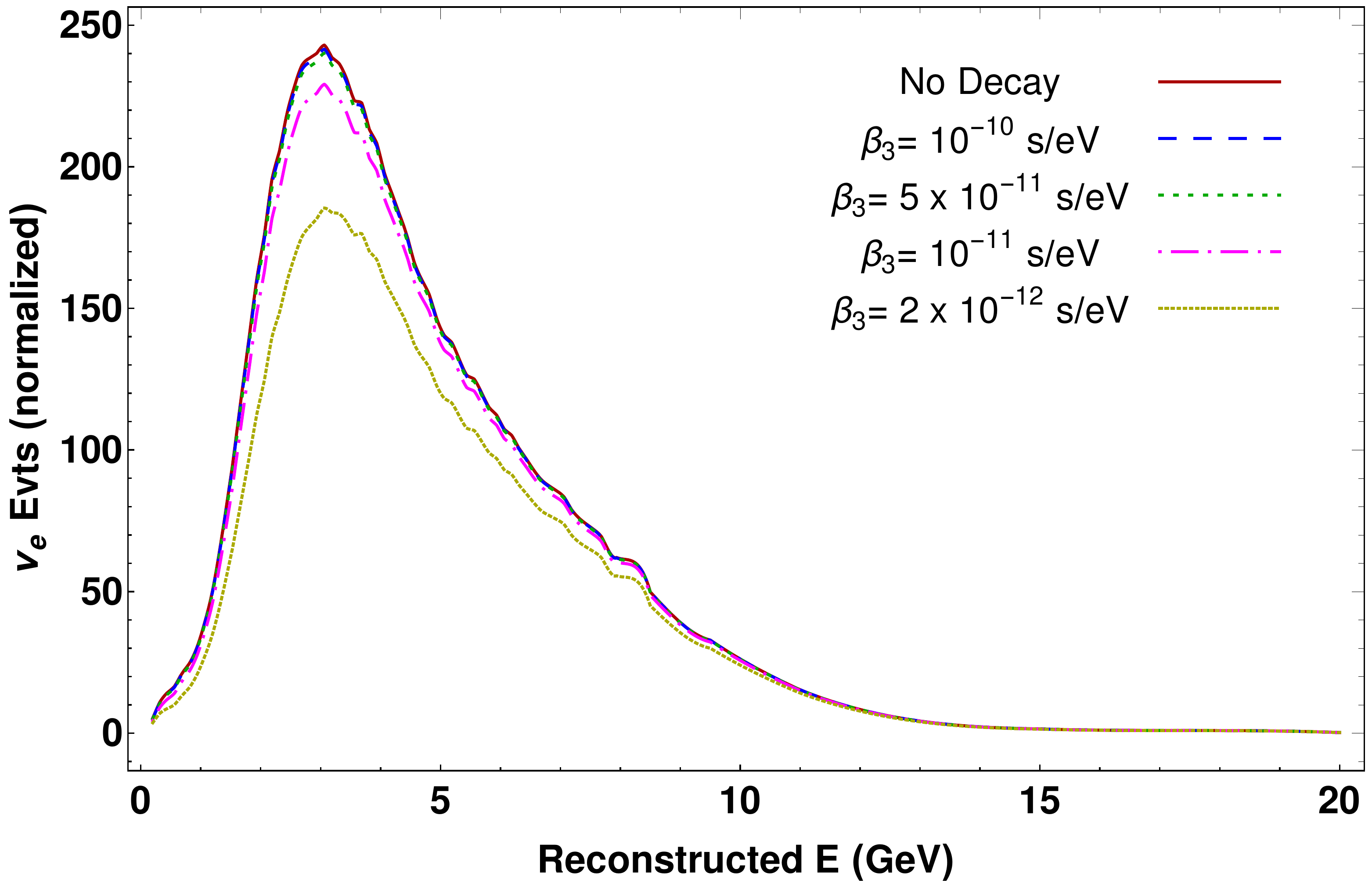} 
\includegraphics[height=6.1cm,width=7.5cm]{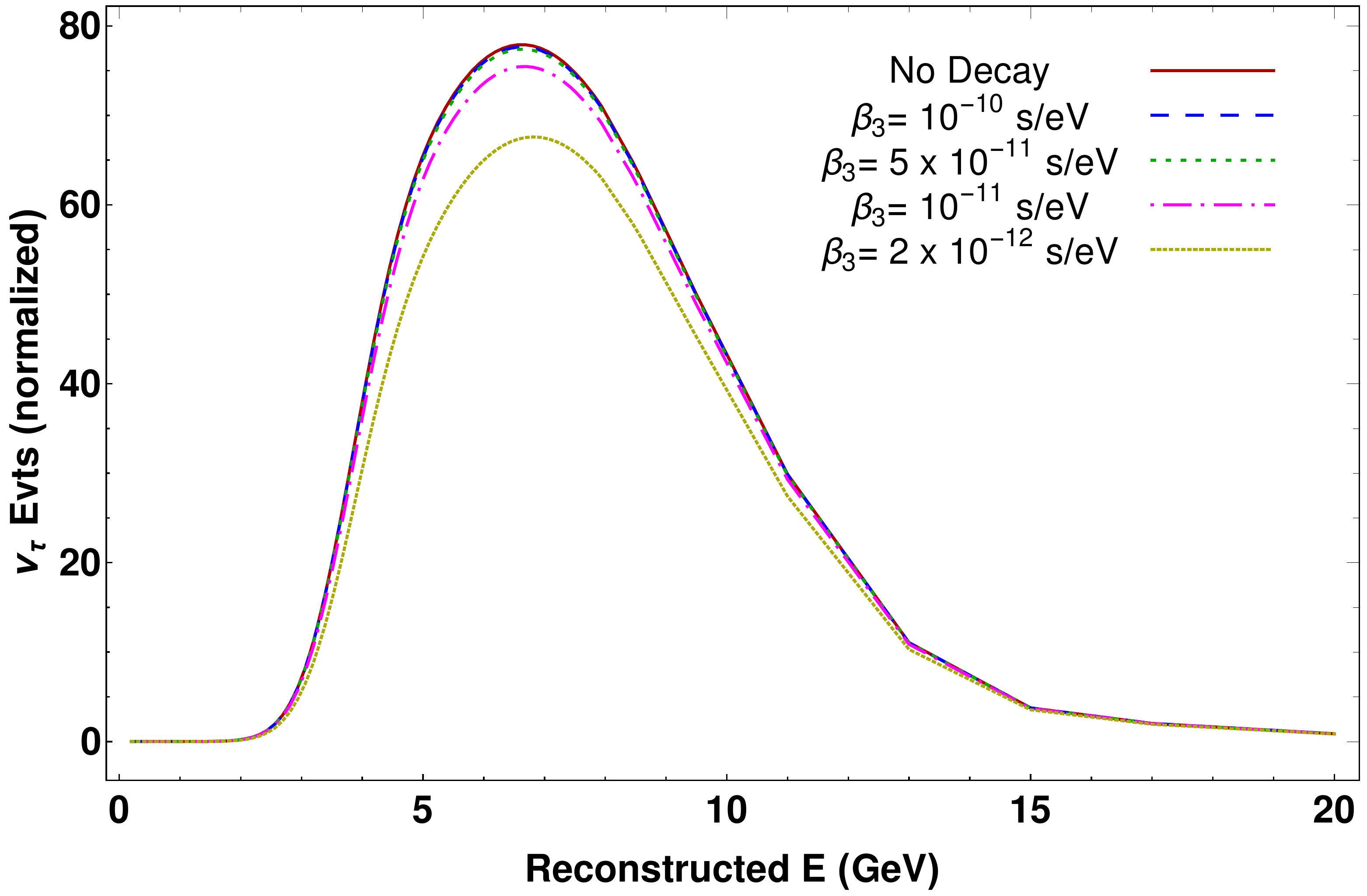} 
\includegraphics[height=6.1cm,width=7.5cm]{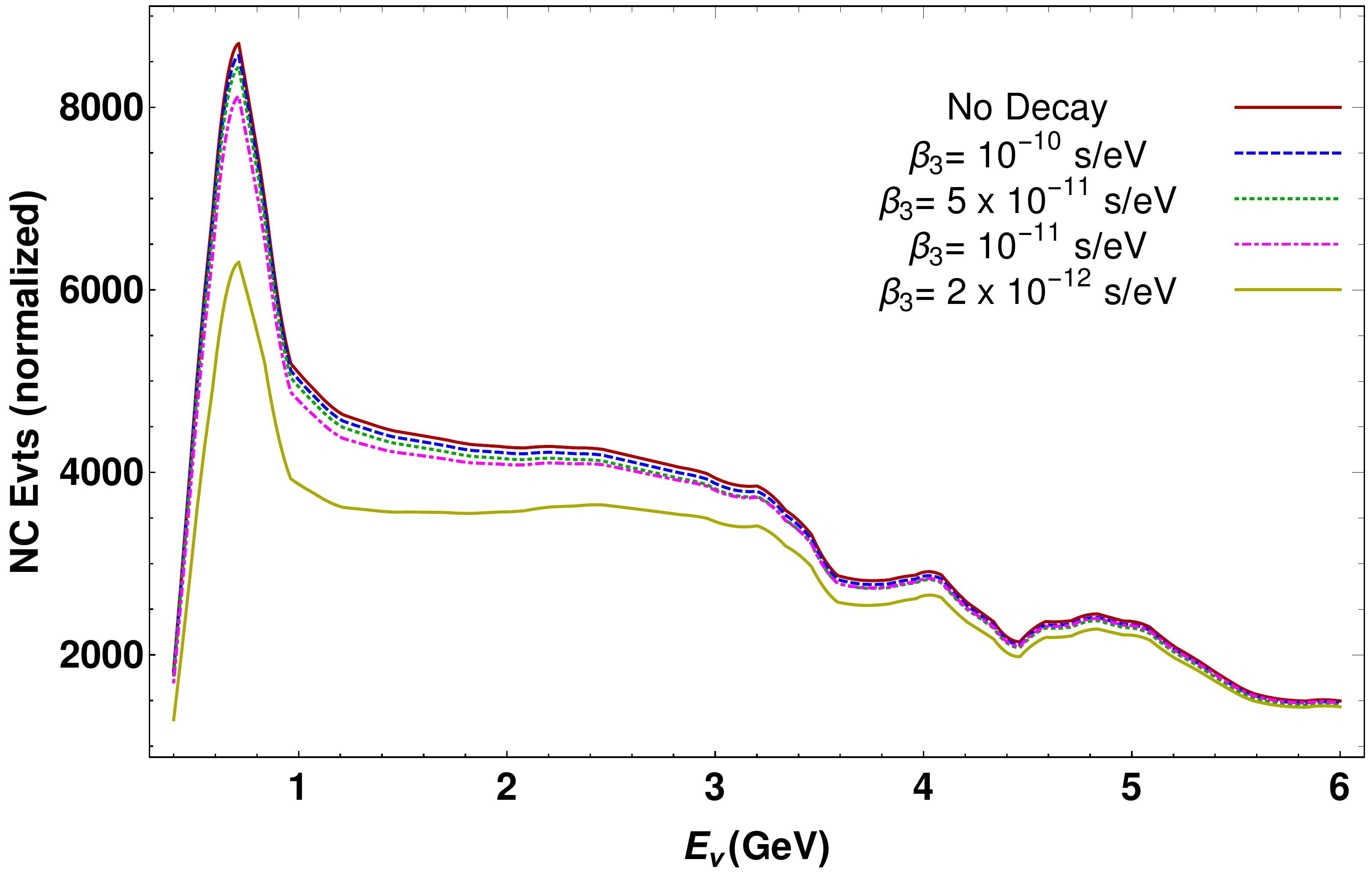} 
\caption{\it Expected $\nu_\mu$, $\nu_\tau$ and $\nu_e$ CC and NC events in DUNE with the optimized flux as a function of the reconstructed neutrino energy for different values of the decay parameter  $\beta_3 = 10^{-10},5\times10^{-11},10^{-11}$ and $\beta_3 = 2\times 10^{-12}~s/eV$. The number of events on the y-axis has been normalized using the variable bin width given by the collaboration in the GLoBES configuration files \cite{Alion:2016uaj}.} 
\label{DUNE_opt}
\end{center}
\end{figure}

\begin{figure}[h]
\begin{center}
\includegraphics[height=6.1cm,width=7.5cm]{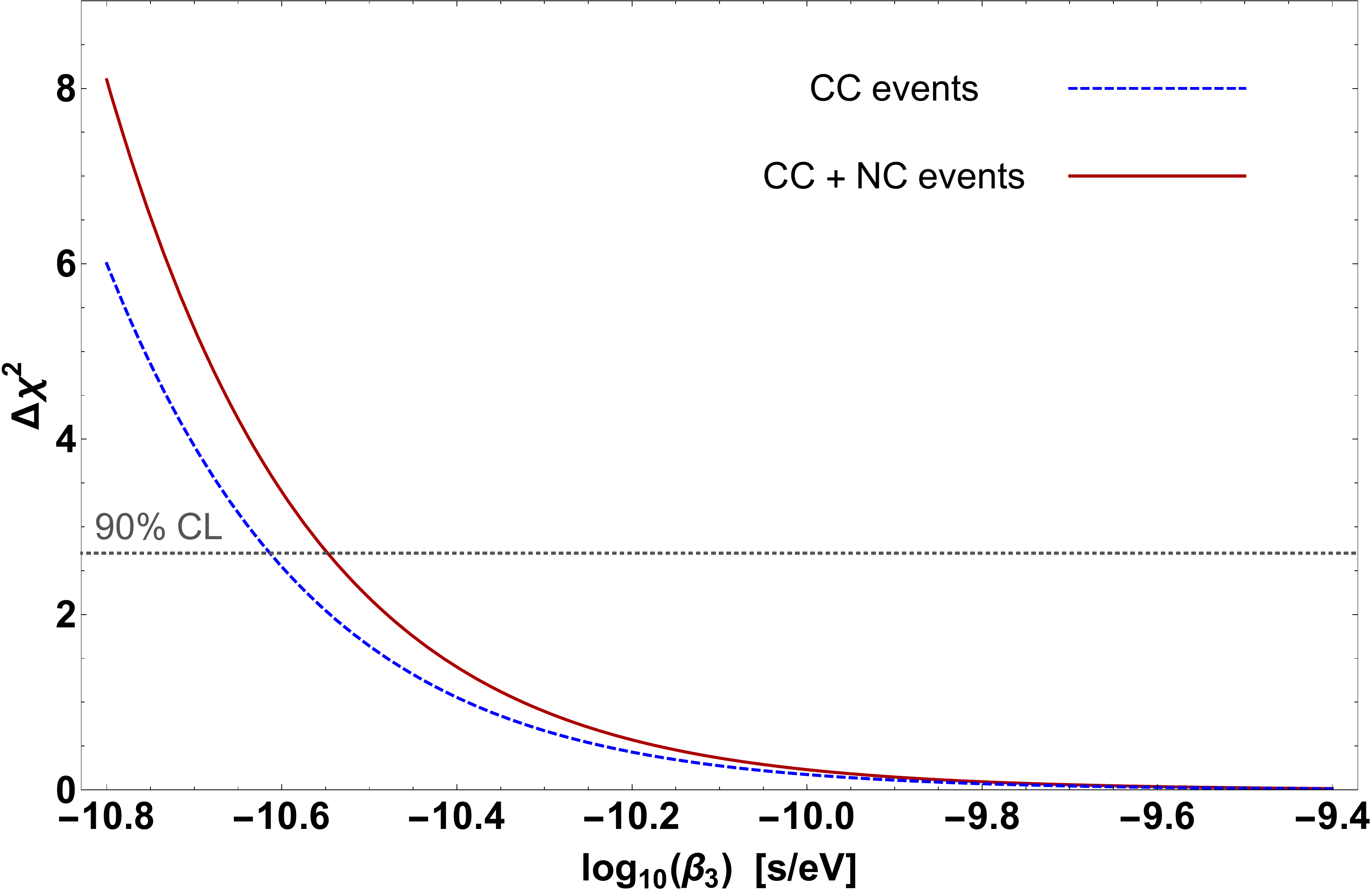} 
\caption{\it Same as Fig.(\ref{alpha_sens}) but for the optimized flux.} 
\label{alpha_sens_opt}
\end{center}
\end{figure}

For the sake of completeness, we show in Fig.(\ref{DUNE_opt}) the expected $\nu_\mu$, $\nu_\tau$ and $\nu_e$ CC and NC events in DUNE with the optimized flux as a  function of the reconstructed neutrino energy for different values of the decay parameter $\beta_3 = 10^{-10},5\times10^{-11},10^{-11}$ and $\beta_3 = 2\times10^{-12}~s/eV$.
This high neutrino energy flux does not allow to study in detail the minimum of the $\nu_\mu \to \nu_\mu$ probability around 2.5 GeV so, compared to Fig.(\ref{DUNE_std}), all the three CC channels appear very similar in shape. 
The effect of the decay parameter is very similar to the one showed for the standard flux, namely a decrease in the number of interactions with a negligible distortion of the shape of the spectra. Despite the larger number of available $\tau$'s, the use of the optimized flux does not help in constraining $\beta_3$ more than the standard flux, Fig(\ref{alpha_sens_opt}), due to the worse performances on the $\nu_\mu \to \nu_e$  and $\nu_\mu \to \nu_\mu$  channels. At 90\% CL we got:
\begin{equation}
\label{bounbsb3_nostro_opt}
\beta_3 > 2.8 \times 10^{-11} \;{\rm s/eV\qquad  (this~~work,~~optimized ~~flux)}  \,.
\end{equation}

\end{document}